\documentclass[fleqn,usenatbib]{mnras}

\usepackage{newtxtext,newtxmath}
\usepackage[T1]{fontenc}

\DeclareRobustCommand{\VAN}[3]{#2}
\let\VANthebibliography\thebibliography
\def\thebibliography{\DeclareRobustCommand{\VAN}[3]{##3}\VANthebibliography}

\usepackage[dvipdfmx]{graphicx}	% Including figure files
\usepackage{amsmath}	% Advanced maths commands
\usepackage{comment}
\title[The role of AGN in the cluster red sequence]{New insights into the role of AGNs in forming the cluster red sequence}

\author[R. Shimakawa et al.]{Rhythm Shimakawa,$^{1,2}$\thanks{E-mail: rhythm.shimakawa@aoni.waseda.jp (RS)}
Jose Manuel P\'erez-Mart\'{\i}nez,$^{3,4,5}$
Yusei Koyama,$^{6}$
Masayuki Tanaka,$^{7}$
Ichi Tanaka,$^{6}$
\newauthor
Tadayuki Kodama,$^{3}$
Nina A. Hatch,$^{8}$
Huub J. A. R\"ottgering,$^{9}$
Helmut Dannerbauer$^{4,5}$
and Jaron D. Kurk$^{10}$
\\
$^{1}$Waseda Institute for Advanced Study (WIAS), Waseda University, 1-21-1, Nishi-Waseda, Shinjuku, Tokyo 169-0051, Japan\\
$^{2}$Center for Data Science, Waseda University, 1-6-1, Nishi-Waseda, Shinjuku, Tokyo 169-0051, Japan\\
$^{3}$Astronomical Institute, Tohoku University, 6-3, Aramaki, Aoba, Sendai, Miyagi 980-8578, Japan\\
$^{4}$Instituto de Astrof\'{\i}sica de Canarias, E-38205 La Laguna, Tenerife, Spain\\
$^{5}$Universidad de la Laguna, Dpto. Astrof\'{\i}sica, E-38206 La Laguna, Tenerife, Spain\\
$^{6}$Subaru Telescope, National Astronomical Observatory of Japan, National Institutes of Natural Sciences, 650 North A'ohoku Place, Hilo, HI 96720, USA\\
$^{7}$National Astronomical Observatory of Japan (NAOJ), National Institutes of Natural Sciences, 2-21-1, Osawa, Mitaka, Tokyo 181-8588, Japan\\
$^{8}$School of Physics and Astronomy, University of Nottingham, University Park, Nottingham NG7 2RD, UK\\
$^{9}$Leiden Observatory, Leiden University, PO Box 9513, NL-2300 RA Leiden, the Netherlands\\
$^{10}$Max-Planck-Institut f\"ur Extraterrestrische Physik, Giessenbachstra{\ss}e 1, D-85748 Garching, Germany
}

\date{Accepted 2024 January 10. Received 2024 January 9; in original form 2023 June 10}

\pubyear{2024}

\begin{document}
\label{firstpage}
\pagerange{\pageref{firstpage}--\pageref{lastpage}}
\maketitle

\begin{abstract}
As a considerable investment of time from various telescope facilities were dedicated toward studying the Spiderweb protocluster at $z=2.2$, it so far remains one of the most extensively studied protocluster.
We report here the latest results in this field, adding a new dimension to previous research on cluster formation at high redshift. 
Previous studies have reported a significant overdensity ($\delta\sim10$) of massive H$\alpha$ (+ [N{\sc ii}]) -emitting galaxies in 3700 comoving Mpc$^3$.
Many of these were previously considered to be dusty, actively star-forming galaxies, given their rest-frame optical and infrared features.
However, this study argues that a third of them are more likely to be ``passively-evolving" galaxies with low-luminosity active galactic nuclei (AGNs) rather than star-forming galaxies, given the multi-wavelength spectral energy distribution (SED) fitting including an AGN component. 
For their SED-based star formation rates to be valid, bulk of their H$\alpha$ + [N{\sc ii}] emission should come from the central AGNs.
This difference in interpretation between this work and past studies, including ours, is particularly supported by the recent deep Chandra/X-ray observation.
Furthermore, we have spectroscopically confirmed a quiescent nature for one of these AGNs, with its multiple stellar absorption lines but also low ionisation emission lines.
This important update provides new insights into the role of AGNs in forming the cluster red sequence observed in the present-day universe.
\end{abstract}


%%%%%%%%%%%%%%%%%%%%%%%%%%%%%%%%%%%%%%%%%%%%%%%%%%

%%%%%%%%%%%%%%%%% BODY OF PAPER %%%%%%%%%%%%%%%%%%

%%%%%%%%%%%%%%%%% INTRODUCTION %%%%%%%%%%%%%%%%%%
\section{Introduction}\label{s1}

Galaxy protoclusters at $z=$ 2--3 are ideal sites for monitoring the transitions from young forming galaxy clusters to evolved clusters in the universe, and the role of the physical drivers in the formation of the tight cluster red sequence thereafter \citep{Bower1992,Kodama2007,Darvish2016,Cai2016,Mei2023}.
During this transition phase, a wide range of physics is involved not only in the protocluster galaxies, but also in the surrounding intergalactic medium, such as cold streams flowing into hot massive haloes, metal-enriched gas recycling, and preheating by energetic feedback \citep{Dekel2009,Valentino2016,ArrigoniBattaia2018,Umehata2019,Daddi2021,Daddi2022,Kooistra2022,Dong2023,Zhang2023}.

Negative (and positive) feedback from active galactic nuclei (AGNs), known as AGN feedback, is thought to be a key mechanism regulating the formation of massive galaxies, particularly at high redshift, and thus the red sequence of clusters (e.g., \citealt{Springel2005,Best2005,Croton2006,Hopkins2006,Schawinski2007,Somerville2008,Kaviraj2011,Alexander2012,Fabian2012,Harrison2012,Schawinski2014,Costa2014,Genzel2014,Shimizu2015,Harrison2017,LeFevre2019,Terrazas2020,Piotrowska2022,Wellons2023,Bluck2023a,Bluck2023b,Byrne2023}).
Galaxy--galaxy interactions are expected to be enhanced in high-$z$ protocluster environments \citep{Okamoto2000,Gottlober2001,Hine2016,Liu2023}, which may trigger AGN activity and hence AGN feedback \citep{Lehmer2009,Lehmer2013,Krishnan2017,Vito2020,Polletta2021,Monson2023}.
Although it is quite challenging to obtain a straightforward proof of AGN feedback in the star formation quenching, the AGN activity has been detected in quiescent galaxies at high redshifts in recent deep observations \citep{Kriek2009,Olsen2013,Marsan2015,Man2016,Gobat2017,Ito2022,Kubo2022,Carnall2023}, providing the role of the AGN activity in the star formation quenching.

In this context, the Spiderweb protocluster, consisting of the radio galaxy, PKS~1138$-$262 at $z=2.16$ \citep{Bolton1979,Roettgering1994,Roettgering1997,vanOjik1995}, and its associated galaxies, is the representative ``maturing" protocluster system that forms a salient red sequence \citep{Kodama2007,Zirm2008,Tanaka2010,Doherty2010,Tanaka2013}.
This system has been intensively studied for years by diverse communities.
The Spiderweb protocluster is the first protocluster at $z>2$ that was specifically targeted and confirmed by \citet{Kurk2000,Pentericci2000}, on the basis of its large rotation measures of the polarised radio emission (\citealt{Carilli1997,Pentericci1997}, see also \citealt{Anderson2022}).
The signal of the thermal Sunyaev-Zeldovich effect has now been detected, suggesting that it is a dynamically active system with M$_{500}=3.46\times10^{13}$ M$_\odot$, before becoming a bona fide cluster seen in the local universe (\citealt{DiMascolo2023}, see also \citealt{Tozzi2022b}). 
Following the initial confirmation of the overdensity traced by Ly$\alpha$ emitting galaxies \citep{Kurk2000,Venemans2007}, significant overdensities of different populations have been repeatedly observed at multiple wavelengths, such as H$\alpha$ emitting galaxies \citep{Kurk2004a,Hatch2011,Koyama2013a,Shimakawa2018}, red sequence galaxies \citep{Kodama2007,Zirm2008,Tanaka2010,Doherty2010,Tanaka2013}, dust-obscured objects \citep{Mayo2012,Koyama2013b,Valtchanov2013,Dannerbauer2014,Zeballos2018}, X-ray sources \citep{Carilli2002,Pentericci2002,Tozzi2022a}, and CO(1--0) emitters (\citealt{Jin2021,Chen2023}, see also \citealt{Emonts2018,Tadaki2019}).
Besides, the kinetic structure of the protocluster system has also been studied through spectroscopic campaigns for associated cluser members \citep{Pentericci2000,Kurk2004b,Croft2005,Shimakawa2014,Shimakawa2015,Jin2021,Perez-Martinez2023}.

Based on the available multi-band data, previous studies have reported more abundant red massive H$\alpha$ emitters (HAEs) with stellar masses M$_\star\gtrsim1\times10^{11}$ solar masses (M$_\odot$) in the Spiderweb protocluster \citep{Hatch2011,Koyama2013a,Koyama2013b,Shimakawa2018}.
These red HAEs were considered to be dusty starbursts before they became bright red sequence galaxies, as seen in the local galaxy clusters.
In fact, they, including the Spiderweb radio galaxy, tend to have irregular and clumpy morphologies in the rest-frame UV images (\citealt{Koyama2013a}, see also \citealt{Stevens2003,Miley2006,Hatch2008}), suggesting that their star formation would be driven by merger events.
However, we need to consider the influence of AGNs on the characterisation of these massive HAEs.
Although we knew this was an important issue \citep{Shimakawa2018,Perez-Martinez2023}, we had overlooked it in our previous work. 
Moreover, the recent extremely deep X-ray observation with Chandra/ACIS-S \citep{Tozzi2022a} has detected more AGN members, and we now have 14 X-ray counterparts out of 30 massive HAEs with M$_\star>2\times10^{10}$ M$_\odot$ in the updated stellar mass estimation (see Section~\ref{s32}), which increases the severity of the issue.

Therefore, this study focuses on the characterisation of 14 massive HAEs \citep{Shimakawa2018} with X-ray counterparts \citep{Tozzi2022a} in the Spiderweb protocluster at $z=2.16$, using multi-wavelength datasets collected from previous studies.
We are primarily interested to revisit their stellar masses and star formation rates (SFRs) in view of AGN contributions.
We also include complementary VLT/KMOS \citep{Perez-Martinez2023} and Keck/MOSFIRE data to validate our conclusions.
In conjunction with the spectral energy distribution (SED) fitting code used in this work \citep{Yang2020}, we assume a flat lambda cold dark matter model with $h=0.693$ and $\Omega_M=0.286$; these values are consistent with those obtained from the WMAP nine-year data \citep{Hinshaw2013}.
We use the \citet{Chabrier2003} stellar initial mass function (IMF) and the AB magnitude system \citep{Oke1983}.
Values for Galactic extinction and dust reddening are respectively based on the \citet{Fitzpatrick1999} extinction curve and \citet{Schlafly2011}.
When we refer to the figures and tables included in this paper, we use capitalised words (e.g., Fig.~1 or Table~1), thereby making them easily distinguishable from those in the literature (e.g., fig.~1 or table~1).

%%%%%%%%%%%%%%%%% DATA AND METHODOLOGY %%%%%%%%%%%%%%%%%%
\section{Data and methodology}\label{s2}

% figure 1
\begin{figure}
\centering
\includegraphics[width=0.95\columnwidth]{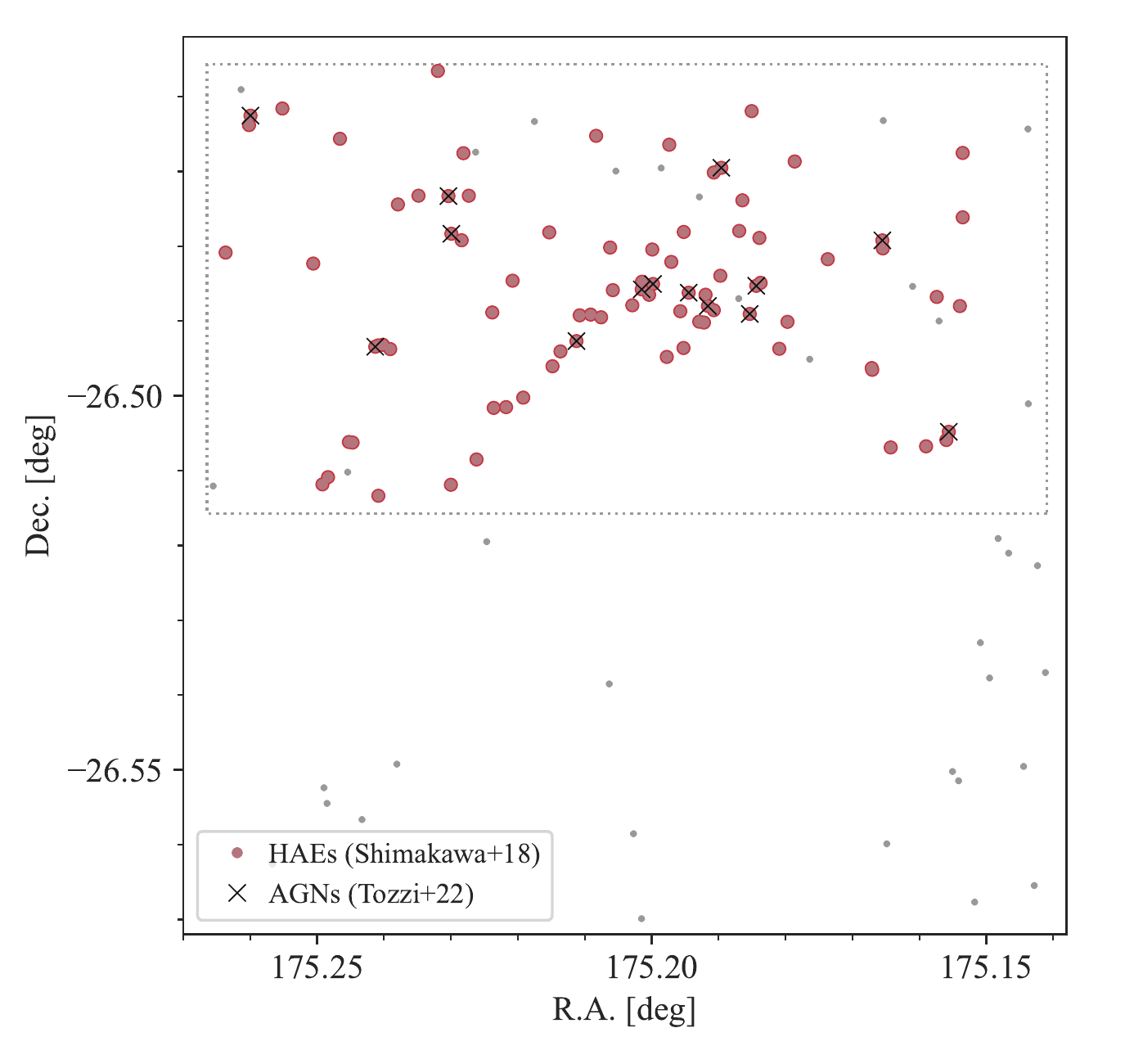}
\caption{
Sky distribution of HAEs \citep{Koyama2013a} and X-ray counterparts \citep{Tozzi2022a}, which are respectively shown by the circle and cross symbols.
The red-filled circles represent the HAE samples in this paper, which are located within the MAHALO-Deep field \citep{Shimakawa2018}, as indicated by the dotted rectangular area.
}
\label{fig1}
\end{figure}

% table 1
\begin{table*}
\centering
\caption{
Catalogue of the X-ray HAEs:
(1) ID numbers in \citet{Shimakawa2018} and (2) \citet{Tozzi2022a}, and (3,4) their coordinates.
(5) Latest spectroscopic redshifts, (6) reference lines of spec-$z$, and (7) these reference papers.
(8) Stellar masses, (9) SFRs, (10) AGN fractions, and (11) AGN type (Sy1 or Sy2) from the {\tt X-CIGALE} code \citep{Yang2020}.
Intrinsic properties of ID=73 (the Spiderweb radio galaxy) may deviate from our measurements as we do not consider the radio component in the SED fitting and also due to its complicated morphology.
ID=77 has no spec-$z$ but is confirmed as Ly$\alpha$ emitter.
Sky coordinates, spectroscopic redshifts, stellar masses and SFRs for all 84 HAE members, including those without X-ray counterparts, are available in Table~\ref{tab4}.
}
\label{tab1}
	\begin{tabular}{ccccccccccc} % four columns, alignment for each
	\hline
	ID & ID$_\mathrm{X}$ & R.A. & Dec. & $z$ & Line & Reference (latest) & M$_\star$ [$10^{10}$ M$_\odot$] & SFR [M$_\star$yr$^{-1}$] & $f_\mathrm{AGN}$ & Type\\
    (1) & (2) & (3) & (4) & (5) & (6) & (7) & (8) & (9) & (10) & (11)\\
	\hline
	14 & 57 & 11:40:37.34 & $-$26:30:17.3 & 2.1684 & H$\alpha$  & \citealt{Perez-Martinez2023} 
       & $19.52\pm6.70$ & $6.92\pm4.67$ & $0.20\pm0.02$ & Sy1\\
	28 & 90 & 11:40:50.70 & $-$26:29:33.6 & 2.1532 & CO         & \citealt{Jin2021}            
       & $7.61\pm3.29$ & $4.82\pm4.70$ & $0.37\pm0.19$ & Sy1\\
	29 & 34 & 11:40:57.91 & $-$26:29:36.3 & 2.1703 & H$\alpha$  & \citealt{Perez-Martinez2023}
       & $7.46\pm5.65$ & $6.18\pm4.79$ & $0.50\pm0.18$ & Sy1\\
	40 & 86 & 11:40:44.48 & $-$26:29:20.6 & 2.1620 & Ly$\alpha$ & \citealt{Croft2005}
       & $9.19\pm2.00$ & $5.33\pm3.43$ & $0.52\pm0.22$ & Sy2\\
	46 & 75 & 11:40:45.98 & $-$26:29:16.7 & 2.1557 & H$\alpha$  & \citealt{Perez-Martinez2023}
       & $2.10\pm3.24$ & $7.87\pm7.66$ & $0.89\pm0.07$ & Sy1\\
	48 & 87 & 11:40:46.67 & $-$26:29:10.3 & 2.1663 & H$\alpha$  & \citealt{Perez-Martinez2023}
       & $18.32\pm2.16$ & $3.53\pm0.45$ & $0.31\pm0.04$ & Sy1\\
	55 & 74 & 11:40:44.25 & $-$26:29:07.0 & 2.1583 & H\&K       & This work (\S\ref{s43})
       & $23.23\pm2.92$ & $2.54\pm1.23$ & $0.40\pm0.21$ & Sy1\\
	58 &  7 & 11:40:47.95 & $-$26:29:06.1 & 2.1568 & H$\alpha$  & \citealt{Perez-Martinez2023}
       & $6.79\pm1.13$ & $41.34\pm15.57$ & $0.75\pm0.08$ & Sy2\\
	68 & 36 & 11:40:39.73 & $-$26:28:45.2 & 2.1620 & Ly$\alpha$ & \citealt{Croft2005}
       & $14.72\pm4.13$ & $35.88\pm30.98$ & $0.39\pm0.17$ & Sy1\\
	71 & 12 & 11:40:55.18 & $-$26:28:42.0 & 2.1630 & H$\alpha$  & \citealt{Perez-Martinez2023}
       & $3.25\pm1.12$ & $3.27\pm3.20$ & $0.51\pm0.14$ & Sy1\\
	73 & 58 & 11:40:48.36 & $-$26:29:08.7 & 2.1618 & CO         & \citealt{Jin2021}
       & $261.7\pm39.4$ & $542.4\pm81.1$ & $0.60\pm0.03$ & Sy2\\
	77 &  9 & 11:40:55.29 & $-$26:28:23.8 & ---    & Ly$\alpha$ & \citealt{Kurk2000}
       & $3.22\pm1.42$ & $5.73\pm4.23$ & $0.19\pm0.19$ & Sy1\\
	83 & 73 & 11:40:45.50 & $-$26:28:10.2 & ---    & ---        &
       & $2.42\pm1.11$ & $5.85\pm6.14$ & $0.40\pm0.27$ & Sy1\\
	95 & 80 & 11:41:02.39 & $-$26:27:45.1 & 2.1510 & H$\alpha$  & \citealt{Perez-Martinez2023}
       & $12.78\pm16.37$ & $65.23\pm29.66$ & $0.62\pm0.04$ & Sy2\\
	\hline
\end{tabular}
\end{table*}

The main focus of this paper is to constrain the stellar masses and star formation rates (SFRs) of 14 X-ray AGNs associated with the Spiderweb protocluster at $z=2.16$ \citep[table~5]{Tozzi2022a} with the Chandra X-ray Observatory \citep{Weisskopf2000}. 
All these have been confirmed in our previous work through H$\alpha$ (+ [N{\sc ii}]) narrow-band imaging \citep{Shimakawa2018} with the Multi-Object InfraRed Camera and Spectrograph (MOIRCS) on the Subaru Telescope \citep{Ichikawa2006,Suzuki2008}. 
We then discuss the role of AGNs in the formation of the red sequence in protoclusters.
We first used 84 HAEs in the Spiderweb protocluster (Fig.~\ref{fig1}), 14 out of which were confirmed as X-ray AGNs (hereafter referred to as X-ray HAEs) by \citet{Tozzi2022a}.
These 84 HAE sources are originally based on 97 narrow-band samples, selected down to the limiting flux of $3\times10^{-17}$ erg~s$^{-1}$cm$^{-2}$ \citep[table~2]{Shimakawa2018}, but with some updates from recent spectroscopic observations \citep{Jin2021,Perez-Martinez2023}.
Among the 84 HAEs, 49 HAEs are identified spectroscopically and the remaining are validated by the $BzK_s$ colour selection \citep{Daddi2004,Daddi2005} to exclude foreground and background emitters (see \S\ref{s213} and \citealt{Shimakawa2018} for details).
Considering previous identifications of $\sim40$ protocluster members that do not overlap with HAEs (e.g., \citealt{Pentericci2000,Tanaka2013,Jin2021}), our HAE sample represents approximately 70\% of the entire protocluster members confirmed to date.
For the X-ray HAEs, 13 out of 14 sources have spec-$z$ confirmations reported in literature \citep{Pentericci2000,Kurk2004b,Croft2005,Tanaka2013,Shimakawa2014,Jin2021,Perez-Martinez2023}.
The detailed information of these X-ray HAEs can be found in Table~\ref{tab1}.

\subsection{Data}\label{s21}

To perform the SED fitting using the {\tt X-CIGALE} (version~2022.1, \citealt{Yang2020}) for the HAE sample, we have collected as much multi-band photometry data as possible from the archive data and literature \citep{Kurk2000,Miley2006,Kodama2007,Seymour2007,Koyama2013a,Valtchanov2013,Dannerbauer2014,Shimakawa2018}, as details are summarised in Table~\ref{tab2}.
Overall, their photometric flux densities and errors were taken from our previous measurements \citep{Shimakawa2017,Shimakawa2018}, which are based on the forced {\tt MAG\_AUTO} photometry using the narrow-band (NB2071) image for source detection with {\tt SExtractor} (version~2.19.5, \citealt{Bertin1996}).
Flux errors were independently measured by the deviations in randomly positioned empty apertures, with corresponding in size to the Kron radii in each photometry to take into account the pixel-to-pixel correlation (see \citealt{Shimakawa2017,Shimakawa2018} for more details).

% table 2
\begin{table}
\centering
\caption{
Photometric datasets used in this work (see references listed in the fifth column for more details).
One should note here that we show the most relevant reference to each data in this work, part of which were independently reduced and analysed across multiple papers.
}
\label{tab2}
	\begin{tabular}{lcccc} % four columns, alignment for each
	\hline
	Telescope/ & Band & $\lambda$ & Depth & Reference\\
	Instrument &      & ($\mu$m)  & (3$\sigma$)\\
	\hline
	Chandra/\\
	ACIS-S & \multicolumn{2}{c}{Hard (2-10~keV)}  & 34.6 & \citealt{Tozzi2022a}\\
           & \multicolumn{2}{c}{Soft (0.5-2~keV)} & 34.0 & $\cdots$\\
	\hline
	VLT/\\
        VIMOS & $U$ & 0.37 & 27.1 & ---\\
	FORS2 & $B$ & 0.43 & 26.8 & \citealt{Kurk2000}\\
	FORS2 & $R$ & 0.65 & 26.2 & $\cdots$\\
	FORS2 & $I$ & 0.79 & 26.4 & $\cdots$\\
	HAWK-I & $Y$   & 1.02 & 26.1 & \citealt{Dannerbauer2014}\\
	HAWK-I & $H$   & 1.62 & 25.1 & $\cdots$\\
	HAWK-I & $K_s$ & 2.15 & 24.8 & $\cdots$\\
	\hline
	Subaru/\\
	S-Cam & $B$        & 0.44 & 26.6 & \citealt{Shimakawa2018}\\
          & $z^\prime$ & 0.91 & 26.4 & \citealt{Koyama2013a}\\
	MOIRCS & $J$   & 1.25 & 24.3 & \citealt{Kodama2007}\\
           & $K_s$ & 2.15 & 24.0 & \citealt{Shimakawa2018}\\
           & NB    & 2.07 & 24.0 & $\cdots$\\
	\hline
	HST/\\
	ACS & F475W & 0.47 & 26.3 & \citealt{Miley2006}\\
        & F814W & 0.81 & 27.0 & $\cdots$\\
	\hline
	Spitzer/\\
        IRAC & ch1 & 3.56 & 21.4 & \citealt{Seymour2007}\\
         & ch2 & 4.50 & 21.6 & $\cdots$\\
	MIPS & 24$\mu$m & 23.6 & 19.9 & \citealt{Koyama2013a}\\
	\hline
	Herschel/\\
    PACS & Green & 100 & 4.5 mJy & \citealt{Seymour2012}\\
         & Red & 160 & 9.0 mJy & $\cdots$ \\
	SPIRE & PSW & 250 & 7.5 mJy & \citealt{Valtchanov2013}\\
          & PMW & 350 & 8.0 mJy & $\cdots$\\
          & PLW & 500 & 9.0 mJy & $\cdots$\\
	\hline
	APEX/\\
    LABOCA & & 870 & 4.5 mJy & \citealt{Dannerbauer2014}\\
	\hline
\end{tabular}
\end{table}

\subsubsection{Chandra X-ray Observatory}\label{s211}

We have adopted the Chandra ACIS-S photometry in the soft (0.5--2.0 keV) and hard (2--10 keV) bands reported by \citet{Tozzi2022a}.
The extremely deep exposure ($\sim700$ ks) reaches the flux limits of $1.3\times10^{-16}$ and $3.9\times10^{-16}$ erg~s$^{-1}$cm$^{-2}$ in the soft and hard bands, respectively, which allowed them to obtain a total of 107 X-ray sources in the range of 2.5 Mpc from the Spiderweb radio galaxy.
Among them, 14 sources have optical counterparts confirmed as HAEs within $\sim1$ arcsec \citep{Shimakawa2018}, including the Spiderweb galaxy itself (Fig.~\ref{fig1}).
For running the {\tt X-CIGALE} SED fitting code \citep{Yang2020}, in this study, their unabsorbed luminosity and errors in the soft and hard bands and photon indexes ($\Gamma$) were taken from table~5 in \citet{Tozzi2022a}.
After transforming the $K$-corrected luminosity to those in the observed frame with $k(z)=(1+z)^{\Gamma-2}$, they were converted to unabsorbed flux densities using the flux converter implemented in the {\tt X-CIGALE} SED fitting code.
We should note that we were forced to adopt observed fluxes and errors for ID = 29 \& 40 (Table~\ref{tab1}), which have been reported as Compton-thick candidates in \citet{Tozzi2022a}.

\subsubsection{Very Large Telescope (VLT)}\label{s212}

The $B$-band data taken from VLT/FORS1 and the $r,i$-band data from VLT/FORS2 were distributed by \citet{Kurk2000,Kurk2004a} via private communication in \citet{Koyama2013a}.
The target region is also covered by the deep near-infrared ($YHK_s$) imaging with VLT/HAWK-I \citep{Dannerbauer2017}.
Photometric measurements for these bands have already been conducted in our previous work \citep{Shimakawa2017,Shimakawa2018}, and thus, we simply adopted the flux densities and errors derived in them.
Furthermore, in this work we have used deep $U$-band data with VLT/VIMOS from the ESO Science Archive (programme ID~383.A-0891).
The $U$-band photometry was performed in the same way as in \citet{Shimakawa2018}, which obtained the forced measurements using the narrow-band (NB2071) image with Subaru/MOIRCS as the reference image.
The point source limiting magnitudes for these datasets are summarised in Table~\ref{tab2}.

\subsubsection{Subaru Telescope}\label{s213}

In addition to the VLT data, the optical ($Bz^\prime$) and near-infrared ($JK_s$ and NB2071) data were obtained with the Suprime-Cam and MOIRCS on the Subaru Telescope, respectively, as a part of the long-running campaign for galaxy clusters and protoclusters at high redshifts, termed the Mapping H-Alpha and Lines of Oxygen with Subaru (MAHALO-Subaru, see also \citealt{Kodama2007,Tanaka2011,Koyama2013a,Shimakawa2018}). 
We employed the same forced measurements in these bands as in \citet{Shimakawa2018}.
The MOIRCS NB2071 filter, with a central wavelength of 2.071 $\mu$m and FWHM $=270$ \AA, is capable of capturing strong emission lines such as the H$\alpha$ line at $z=2.155\pm0.020$ and the [O{\sc iii}]$\lambda5008$ line at $z=3.135\pm0.026$.

Based on these data, the narrow-band selection ($K_s-$~NB2071) and the $Bz^\prime K_s$ colour--colour diagram were conducted for the HAE selection at $z=2.16$, as described in more detail by \citet{Shimakawa2018}.
The narrow-band selection in our previous work detected 97 narrow-band emitters above the three sigma confidence level, of which 36, 32, and 13 emitters were further selected as spec-$z$ (or dual Ly$\alpha$ and H$\alpha$) members, $Bz^\prime K_s$ colour-selected members, and member candidates that could not be rejected by the colours, respectively.
The remaining 16 narrow-band emitters were defined as foreground or background emitters.
The expected contamination rate for the 13 member candidates is $\sim10$\%, given the follow-up spectroscopy \citep{Shimakawa2018}.
Subsequently, the number of spec-$z$ confirmations was increased to $N=49$ by the ATCA CO(1-0) observation \citep{Jin2021} and the VLT/KMOS $K$-band spectroscopy \citep{Perez-Martinez2023}.
As a result, we now have 49, 23, and 12 HAEs with spec-$z$ confirmations, colour validations, and candidates, respectively; hence, a total of 84 ($=49+23+12$) protocluster members of HAEs in this paper.
However, it should be noted that this paper solely focuses on 14 X-ray HAEs and the remaining 70 non-X-ray HAEs are used only as a reference sample but summarised in Appendix~\ref{a1} and Table~\ref{tab4},~\ref{tab5}.

\subsubsection{Hubble Space Telescope}\label{s214}

We have included the Hubble/ACS (F475W, F814W) data from the Hubble Legacy Archive. 
The Hubble/ACS data were first used by \citet{Stevens2003} and have been studied repeatedly in several papers \citep{Miley2006,Hatch2008,Koyama2013a,Naufal2023}.
In this work, we do not use the original high-resolution images, but adopt the PSF-convolved forced measurements of \citet{Shimakawa2018}.
The obtained flux densities are consistent with those in the similar bands from the ground-based telescopes, as seen in Section~\ref{s31}.

\subsubsection{Spitzer Space Telescope}\label{s215}

In addition, we used the Spitzer/IRAC 3.6 (ch1) and 4.5 (ch2) $\mu$m bands from the NASA/IPAC Infrared Science Archive (see also \citealt{Seymour2007}), which cover 92\% of the HAE sample (or $\sim80$\% of the survey area, Fig.~\ref{fig1}), and 13 out of 14 X-ray HAEs in this work (only missing ID=95 in the ch1 band, Table~\ref{tab1}).
As described in \citet{Shimakawa2018}, we used the Post-BCD (PBCD) products from the Spitzer data archive library.
However, we modified the IRAC photometry from the literature based on the forced measurements on the narrow-band (NB2071) image, as for the other photometric bands, instead of the independent measurements performed in the previous work.
Here, we obtained a fixed aperture photometry with a diameter of 4 arcsec, and then performed the aperture corrections by multiplying by 1.20 and 1.22 in the ch1 and ch2 bands, respectively.
This revision somehow helps us constrain the stellar masses of the X-ray HAEs and the upper limits for the lower-mass HAEs.

We also adopted the Spitzer/MIPS 24$\mu$m source photometry based on the reduced data provided by \citet[see also \citealt{Mayo2012}]{Koyama2013a}.
As in the case of the IRAC photometric estimates, we re-measured flux densities using fixed aperture photometry, but with a diameter of 7 arcsec, and then applied the aperture correction by multiplying by a factor of 2.8.
Due to the large seeing size (6 arcsec), we overestimated the MIPS flux densities in some cases that the targets have close neighbours.
However, we stress that the addition of MIPS photometry does not make a significant difference to our conclusions.

\subsubsection{Far-infrared and Submillimetre Data}\label{s216}

Furthermore, we made the best possible effort to constrain the upper limits of SFRs by including public source photometry and limiting fluxes from \citet[table~2]{Dannerbauer2017}, which are based on far-infrared to submillimetre (submm) observations with the Herschel Space Telescope and the Atacama Pathfinder Experiment (APEX).
However, the current shallow far-infrared and submm data show slight effect on the SED fitting results.
We have used flux densities and errors in the Herschel/PACS 100 and 160 $\mu$m bands \citep{Seymour2012}, Herschel/SPIRE 250, 350, 500 $\mu$m bands \citep{Valtchanov2013}, and APEX/LABOCA 870 $\mu$m band for the HAEs with submm counterparts \citep{Dannerbauer2017}, and the $3\sigma$ limits to constrain the upper limits for the remaining HAEs (Table~\ref{tab2}).
For the 14 X-ray HAEs, only two of the most massive systems (ID=73 and 95), including the Spiderweb radio galaxy, are reliably detected by \citet[DKB07 and DKB16]{Dannerbauer2017}.

In addition, we adopted the recent 870 $\mu$m flux measurement in ALMA Band~7 for 4 X-ray HAEs (ID=14, 46, 48, 71 in Table~\ref{tab1}), where the details of the data processing and photometry are explained by Koyama et al. (in preparation).
We confirmed that the ALMA Band~7 data have only a small impact on the derivation of their stellar masses and SFRs in the SED fitting and hence do not affect the conclusions of this paper as in the case of the other infrared and submm data.
% However, such 870 $\mu$m detections provide very useful information to discuss the validation of the main results in this paper (Section~\ref{s32}).

\subsection{SED modelling}\label{s22}

In order to obtain physical quantities for AGN host galaxies, it is very important to perform a SED fitting taking into account the AGN component.
We therefore ran the SED fitting code, {\tt X-CIGALE} (version 2022.1, \citealt{Boquien2019,Yang2020,Yang2022}), to characterise the X-ray HAEs, including the other HAEs for reference.
The {\tt X-CIGALE} is excellent for the X-ray AGNs like our targets, as it not only covers the ultraviolet to radio regimes, but also has an X-ray module.
The photometric fluxes and errors in the multi-wavelength bands were taken from our parevious measurements \citep{Shimakawa2017,Shimakawa2018}, as described in the last subsection.

For the SED fitting, this paper generally follows the module selection in the original paper \citep{Yang2020}.
Specifically, we used the stellar templates of \citet{Bruzual2003} with a fixed stellar metallicity of $Z=0.004$ (0.2 Z$_\odot$) and the \citet{Chabrier2003} IMF, assuming a delayed exponentially declining star formation history (SFH) $\propto t\cdot\exp(-t/\tau)$.
We refer to \citet{Pearson2017} for the selection of stellar population models in the SFH: the values of $t$ and $\tau$ are allowed to be {\tt age\_main} $=[0.5, 1, 2, 3]$ and {\tt tau\_main} $=[0.1, 0.3, 0.5, 1, 3, 5]$ Gyrs in the main stellar populations and {\tt age\_burst} $=[0.001, 0.01, 0.03, 0.1, 0.2]$ and {\tt tau\_burst} $=10^6$ Gyrs in the late burst, respectively.
Here, the late burst with $\tau=10^6$ Gyrs approximates constant star formation.
The mass fraction of the late burst is allowed to be [0.001, 0.01, 0.1, 0.2].
The selection of stellar populations in the SFH is most critical for the derivation of stellar masses and SFRs, and moderately produces the systematic difference.
Particularly, stellar masses (SFRs) of some young, low-mass HAEs, which are beyond the scope of this paper, may be over-(under-)estimated (Appendix~\ref{a1}).
However, it should be noted that this does not change the obtained trends and hence the main conclusion of this paper.
An additional explanation of the coherence of the stellar mass estimation between this work and the previous study \citep{Shimakawa2018} is provided in Appendix~\ref{a1}.

% table 3
\begin{table}
\centering
\caption{
Fitting parameters of AGN and X-ray models adopted by this work in the X-CIGALE code \citep{Yang2020}.}
\label{tab3}
	\begin{tabular}{lc} % four columns, alignment for each
	\hline
	Module/   & Values\\
	Parameter &\\
	\hline
	SKIRTOR2016/\\
	Torus optical depth at 9.7 $\mu$m          & 7\\
	Torus density radial parameter             & 1\\
	Torus density angular parameter            & 1\\
	Angle between the equatorial plan and edge of the torus & 40\\
	Ratio of the max to min radii of the torus & 20\\
	Viewing angle                              & 30, 70\\
	AGN fraction in total IR luminosity        & 0.01--0.99\\
	Extinction law of polar dust               & SMC\\
	$E(B-V)$ of polar dust                     & 0.01--1\\
	Temperature of polar dust                  & 100\\
	Emissivity of polar dust                   & 1.6\\
	\hline
	XRAY/\\
	AGN photon index                           & 1.8\\
	Max deviation from $\alpha_\mathrm{ox}$--$L_\mathrm{2500}$ relation & 0.2\\
	LMXB photon index                          & 1.56\\
	HMXB photon index                          & 2.0\\
	\hline
\end{tabular}
\end{table}

The model fit also includes nebular templates with an ionisation parameter ($\log U=-2$) and $Z_\mathrm{gas}=0.02$ based on \citet{Inoue2011} computed from the {\tt Cloudy} 0.8.00 photoionization code \citep{Ferland2017}.
We have adopted the \citet{Calzetti2000} attenuation law with $E(B-V)=$ 0.01--1.0 mag and dust emission from \citet{Draine2014} without changing the default model parameters.
In addition, the SED models include AGN components with AGN fractions $f_\mathrm{AGN}=$ 0.01--0.99, which is defined by the fraction of AGN infrared luminosity to the total infrared luminosity \citep{Yang2020}.
X-ray emissivity and AGN models are based on \citet{Stalevski2016}, where we selected the same model parameters (Table~\ref{tab3}) as in \citet[table~3]{Yang2020} except for $E(B-V)$, which is allowed to be 0.01--1.0 mag by considering high column densities of some X-ray HAEs (\citealt{Tozzi2022a}, and see also Section~\ref{s42}).
It should be noted that the model selection adopted here may not be the best choice for our sample, and these parameters have rather been selected as a compromise because of the limited data availability (see also Section~\ref{s31}).
We also include the effect of IGM absorption based on \citet{Meiksin2006}.
Consequently, a total of 15,396,480 models per redshift were adopted to conduct the SED fitting for the X-ray HAEs.
It took about one hour to run the SED fitting in the environment of $28\times3.3$ GHz Intel CPU.

%%%%%%%%%%%%%%%%% RESULTS %%%%%%%%%%%%%%%%%%
\section{Results}\label{s3}

\subsection{Result of SED fitting}\label{s31}

% figure 2
\begin{figure*}
\centering
\includegraphics[width=0.94\textwidth]{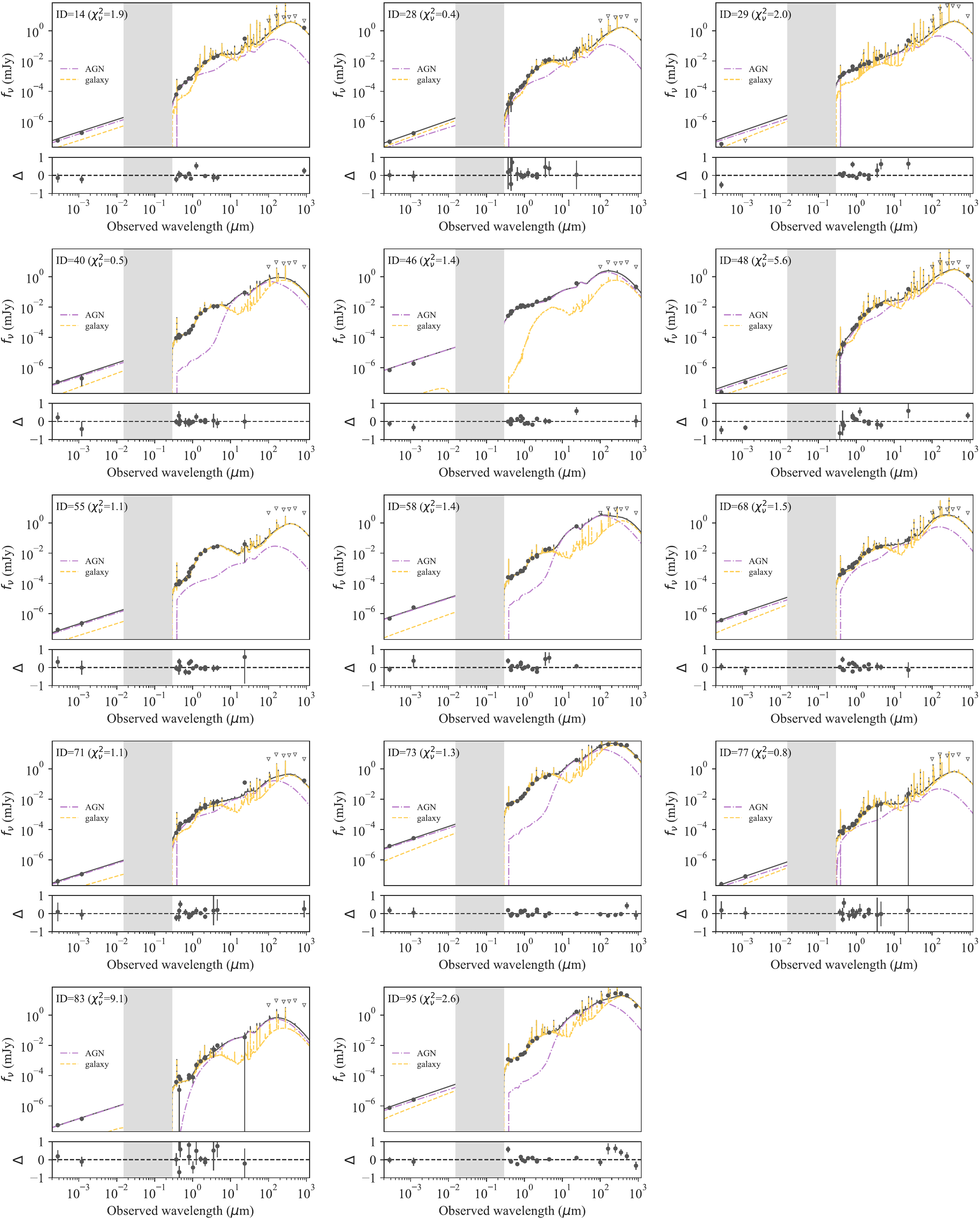}
\caption{
The best-fit SED spectra (black lines on the upper panels) and normalised residual errors $\Delta$ (black circles on the lower panels) of the X-ray HAEs obtained by the {\tt X-CIGALE} code.
The relative flux residuals $\Delta$ are defined by $\Delta\equiv(f_\mathrm{\nu,in}-f_\mathrm{\nu,out})/f_\mathrm{\nu,out}$, where $f_\mathrm{\nu,in}$ and $f_\mathrm{\nu,out}$ are observed flux densities and the best-fit flux densities in the broad-bands, respectively.
The identification number and the best-fit reduced chi-squares are denoted at upper left in each panel.
The filled black circles show observed flux densities in the filter bands for each target, while the opened inverted triangles are the upper limits adopted for non-detection in the infrared and/or submillimetre bands (Section~\ref{s215} and \ref{s216}).
The magenta and yellow lines on the upper panels depict AGN and galaxy components, respectively.
We conveniently masked an unused wavelength range from the rest-frame 5 nm (corresponding to the limit of the X-ray module) to the Lyman limit by the grey regions (see Section~\ref{s31}).
}
\label{fig2}
\end{figure*}

The best-fit SEDs of the X-ray HAE samples are shown in Figure~\ref{fig2}, indicating that the SED fitting worked well with the best reduced chi-squares $\chi^2_\nu<5$, except for the two X-ray HAEs (ID=48 and 83 with $\chi^2_\nu=$ 5--10).
The derived stellar masses, SFRs, AGN fractions, and AGN types and their associated errors are summarised in Table~\ref{tab1}.
Overall, the SED fitting with the AGN components and the X-ray modules help to decrease the reduced chi-squares by a factor of 1.5 on average, and makes the host galaxy spectra redder owing to AGN contributions in the rest-frame ultraviolet.
For example, the median contribution at $\lambda_\mathrm{rest}=2800$~\AA\ is as much as 48\%, while it is only approximately 18\% in the rest-frame optical band ($\lambda_\mathrm{rest}=6500$~\AA).
In the most cases, host galaxy components are more dominant than AGNs in the rest-frame optical, except for two X-ray HAEs (ID=29 and 46, Figure~\ref{fig2}).
On the other hand, obtained AGN fractions in the infrared luminosity ($f_\mathrm{AGN}$) exceed 50\% in six X-ray HAEs (Table~\ref{tab1}).
Particularly, the best-fit stellar mass of one of the X-ray HAEs (ID=46) significantly decreases by a factor of 3.5 due to a high AGN fraction of $f_\mathrm{AGN}=0.89$ when the AGN component is included. 
However, we note that it largely depends on the SED modelling.
It should be noted that there may be larger uncertainties beyond the inferred values in the Spiderweb radio galaxy (ID=73) owing to its extended and complex morphology.
In addition, we have confirmed that SED-based SFRs of non-X-ray HAEs are self-consistent with their H$\alpha$-based SFRs based on the narrow-band fluxes of \citet{Shimakawa2018}, as shown in Appendix~\ref{a1}.

Figure~\ref{fig2} also suggests that the best-fit SEDs broadly agree with the X-ray observation in the most cases; however, there is still room for improvement.
Whereas, determining more complex parameters with a limited sample size and only the two X-ray bands is difficult and may need investigations from a more fundamental aspect, which is beyond of the scope of this paper.
In addition, the number of the SED models in this work (15,396,480 per redshift) is close to the cache size in our system environment.
Detailed analyses of local relics of high-redshift AGNs will be helpful to optimise the parameter selection and to gain a more precise understanding of characteristics of such X-ray HAEs at $z>2$.
Besides, one should note that the {\tt X-CIGALE} covers a wavelength range up to the rest-frame 5 nm in the X-ray module \citep{Yang2020} and its IGM absorption model does not consider the Lyman Limit Systems.
However, our photometric data do not cover the rest-frame 5 nm to the Lyman limit, and thus these issues do not affect our SED fitting results practically (Fig.~\ref{fig2}).

As mentioned in Section~\ref{s21}, we confirmed that there is no significant impact on observational trends (see Section~\ref{s32}) even without infrared and submm data at $\lambda>20$ $\mu$m.
Nevertheless, it can be seen that their flux densities and the upper limits, are broadly consistent with the best-fit SEDs (Fig.~\ref{fig2}).
There are significant flux excesses in the MIPS $24\mu$m band in some cases, which seem to be caused by blending issue, given the limited seeing size at FWHM of 6 arcsec.
The JWST Mid Infrared Instrument (MIRI, \citealt{Rieke2015}) will be able to provide more reliable photometry and spectra with much better spatial resolution (FWHM $\sim0.8$ arcsec) in the similar infrared bands.
Additionally, combined high-resolution data from the upcoming JWST/NIRCam run \citep{Dannerbauer2021} and the existing HST/ACS data \citep{Miley2006} may help to achieve the galaxy--AGN decomposition in a more immediate manner, especially for the X-ray HAEs with high AGN fractions (e.g., see \citealt{Ding2022}).

\subsection{Star-forming main sequence}\label{s32}

% figure 4
\begin{figure*}
\centering
\includegraphics[width=0.9\textwidth]{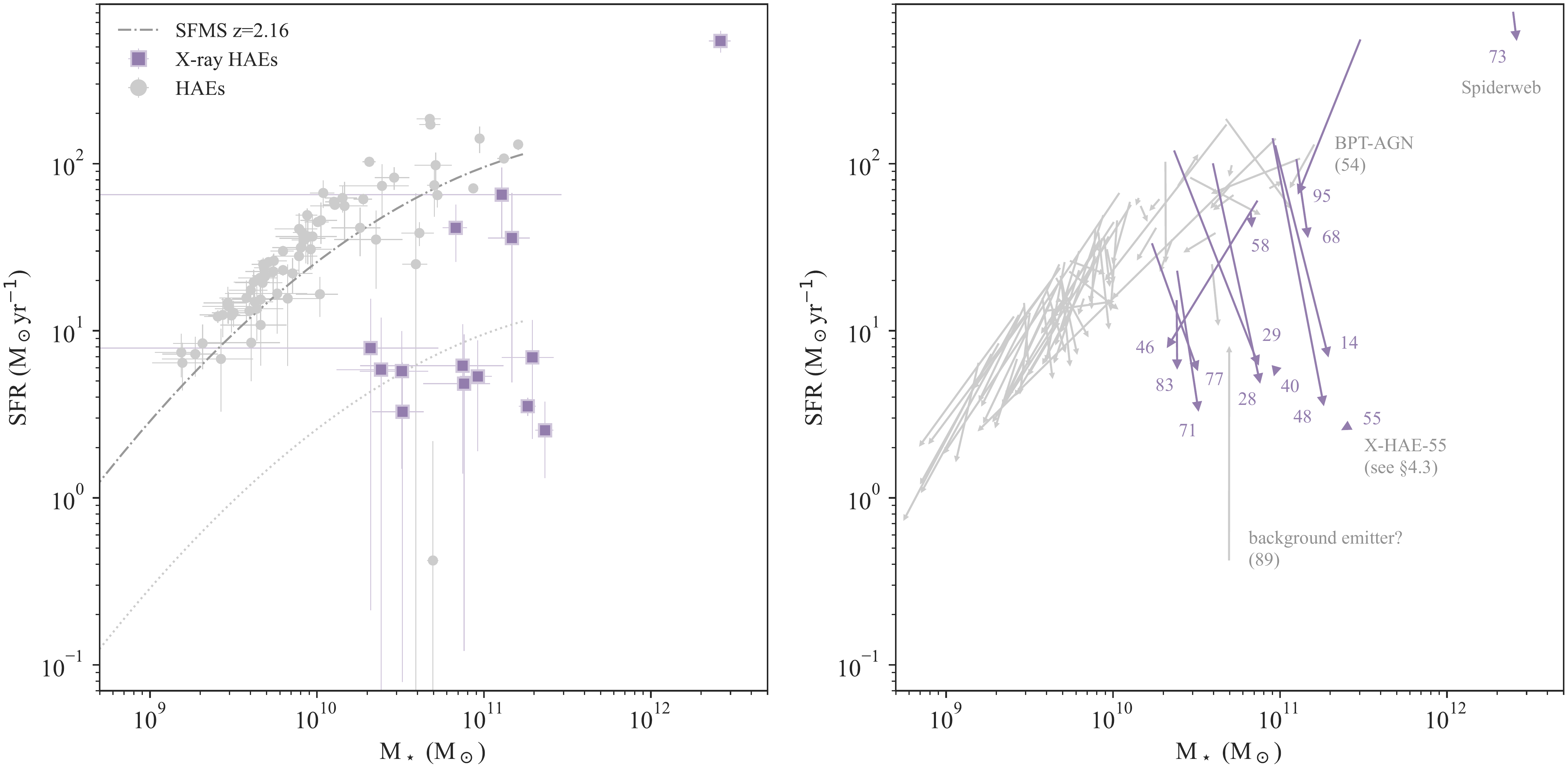}
\caption{
The left panel shows SED-based SFRs versus stellar masses (M$_\star$) of 84 HAEs in the Spiderweb protocluster at $z=2.2$ \citep{Shimakawa2018}.
The purple and grey symbols represent HAEs with ($N=14$) and without ($N=70$) X-ray counterparts \citep{Tozzi2022a}, respectively.
We here do not consider the AGN components in the SED fitting for the latter case. 
The black dot-dash curve represents the star-forming main sequence at $z=2.16$ from \citet{Popesso2023} with the IMF correction from \citet{Kroupa2001} to \citet{Chabrier2003}, and its $\times0.1$ relation is depicted by the grey dotted curve.
On the right panel, arrow symbols depict how these values change on the SFR--M$_\star$ plane when assuming the AGN components with AGN fractions $f_\mathrm{AGN}\geq0.01$ in the SED fitting (see text), each of which is attached with the identification number in Table~\ref{tab1}.
Additional comments are included in the figure for the sake of clarity.
}
\label{fig3}
\end{figure*}

It is well known that there is a tight relationship between star formation rates and stellar masses of galaxies termed star-forming main sequence \citep{Brinchmann2004,Noeske2007,Daddi2007,Elbaz2007,Salim2007,Whitaker2012,Speagle2014,Lee2015,Renzini2015,Tomczak2016,Tacchella2016,ForsterSchreiber2020,Popesso2023}.
The resulting SFRs and stellar masses are represented in Figure~\ref{fig3}, indicating that 10 out of 30 massive HAEs with M$_\star>2\times10^{10}$ M$_\odot$ are located significantly ($\sim1$ dex) below the star-forming main sequence formed by non-X-ray HAEs (and e.g., \citealt{Popesso2023}), despite of the detection of H$\alpha+$ [N{\sc ii}] line emissions.
Most of them (9/10) are HAEs with low X-ray luminosity ($L_X<4\times10^{43}$ erg~s$^{-1}$ in the hard band), where one of the X-ray HAEs (ID=46) with low SFR has not been counted due to its substantial margin of error in the stellar mass.
Figure~\ref{fig3} also indicates approximately half of massive HAEs with M$_\star>2\times10^{10}$ M$_\odot$ host X-ray AGNs ($\sim14/30$), which yields the AGN fraction of $0.42\pm0.03$ when including three additional quiescent galaxies with spectroscopic identifications \citep{Tanaka2013}.

We show how the addition of the AGN templates ($f_\mathrm{AGN}\geq0.01$) and the X-ray module affects their stellar mass and SFR measurements, as shown in the right panel of Figure~\ref{fig3}.
It can be seen that the addition of the AGN components has a significant effect on quiescent properties of the X-ray HAEs, except for the two X-ray HAEs (ID = 40 \& 55) that remain quiescent irrespective of the AGN models.
On the other hand, most of massive star-forming HAEs without X-ray counterparts in \citet{Tozzi2022a} are significantly shifted from the star-forming main sequence to the passive sequence, assuming that they have AGN components with an AGN fraction of $f_\mathrm{AGN}\geq0.01$.
Here, their X-ray flux densities are constrained by the upper limits (Table~\ref{tab2}) but multiplied by 1.5 for the soft band taking account of typical X-ray absorption in the X-ray HAEs.
The result suggests that the recent deep Chandra X-ray observation \citep{Tozzi2022a} plays a rather important role in updating the SFR measurements of these X-ray HAEs, as well as in revealing such intriguing properties of AGN host galaxies in the Spiderweb protocluster.

With regard to the non-X-ray HAEs, we observe only one non-X-ray HAE (1/10) lying well below the star-forming main sequence (ID=89 in \citealt{Shimakawa2018}) regardless of whether we assume that they host AGNs with the AGN fraction $f_\mathrm{AGN}>0.01$ or not (Fig.~\ref{fig3}). 
The HAE ID=89 could be a dusty [O{\sc iii}] emitter at $z\sim3.1$ or [O{\sc ii}] emitter at $z=4.6$, given no clear detection in the $UBr$-bands.
We also note that the most massive non-X-ray HAE (ID=54) would be an AGN according to the BPT diagram \citep{Shimakawa2015,Perez-Martinez2023}, which further supports the very high AGN fraction in the Spiderweb protocluster.

Importantly, as noted in Section~\ref{s22}, we cannot completely deny the possibility that the obtained trend could be model dependent, although we confirmed that the trend itself does not change even if we chose different model settings with best efforts.
Direct evidence is highly desirable to ensure star formation quenching in these AGN host galaxies.
At present, there is no sufficient observational constraint to determine the nature of passive evolution for the entire X-ray HAE sample, although we show a spectroscopic identification of the star formation quenching for one of the X-ray HAEs (ID=55) in Section~\ref{s43}.

%%%%%%%%%%%%%%%%% DISCUSSION %%%%%%%%%%%%%%%%%%
\section{Discussion}\label{s4}

\subsection{Protocluster versus field}\label{s41}

We found that a third of 30 massive HAEs with M$_\star>2\times10^{10}$ M$_\odot$ in the Spiderweb protocluster at $z=2.16$ have unexpectedly low SFRs, despite their high H$\alpha$ + [N{\sc ii}] fluxes $>4\times10^{-17}$ erg~s$^{-1}$cm$^{-2}$ \citep{Shimakawa2018}.
Because the less active HAEs host low luminosity AGNs ($L_X\lesssim4\times10^{43}$ erg~s$^{-1}$) detected by the recent deep Chandra X-ray observation \citep{Tozzi2022a}, the bulk of their H$\alpha$ (+ [N{\sc ii}]) emission would be driven by AGNs rather than star formation (see also Section~\ref{s42}).
Our result suggests that the formation process of the red sequence galaxies is well underway in the Spiderweb protocluster faster than previously thought \citep{Koyama2013a,Shimakawa2018}.
Considering recent identifications of an intracluster medium (ICM) with the deep X-ray and submm observations \citep{Tozzi2022b,DiMascolo2023}, the state of the Spiderweb protocluster seems to be very close to the bona fide clusters observed at $z\lesssim2$.
In this context, some passive AGNs in the protocluster centre might have contributed to preheating of protocluster ICM \citep{Kooistra2022,Dong2023}, although high-resolution IGM tomography is required to test the preheating scenario in greater detail.
Additionally, the result suggests that AGNs may be involved in their star formation quenching. 
However, it is premature to get to the conclusion at this point, without identifying physical associations between AGNs and host galaxies.
The remainder of this subsection discusses whether such an intriguing result is a peculiar trend in high-density environments or more ubiquitous throughout the universe at high redshifts.

To discuss the environmental dependence, we used the X-ray selected AGNs at $z\sim2$ in the COSMOS field \citep{Scoville2007} and the GOODS-S field \citep{Giavalisco2004}, which are based on the Chandra-COSMOS Legacy survey \citep{Civano2016} and the COSMOS2020 catalogue \citep{Weaver2022}, and the Chandra Deep Field-South Survey \citep{Luo2017,Liu2017} and the 3D-HST catalogue \citep{Skelton2014}, respectively. 
In the COSMOS field, a total of 51 sources with spectroscopic redshifts of $z=$ 1.5--2.5 over $\sim2.2$ deg$^2$ are selected from \citet{Suh2020} with X-ray fluxes, down to $3.7\times10^{-16}$ erg~s$^{-1}$cm$^{-2}$ in the soft band and $1.5\times10^{-15}$ erg~s$^{-1}$cm$^{-2}$ in the hard band. 
They are respectively higher than the flux limits of $1.3\times10^{-16}$ erg~s$^{-1}$cm$^{-2}$ and $3.9\times10^{-16}$ erg~s$^{-1}$cm$^{-2}$ in the Spiderweb protocluster \citep{Tozzi2022a}.
Therefore, a clear sampling bias can be noted in the sense that we are missing X-ray faint AGNs in the COSMOS field.
Specifically, we adopted source photometry in the CLAUDS $U$-band \citep{Sawicki2019}, $grizy$ bands from the HSC-SSP PDR2 \cite{Aihara2018,Aihara2019}, $BgVrizz^\prime$ and several medium bands from the Subaru Suprime-Cam data \citep{Taniguchi2007,Taniguchi2015}, $YJHK_s$ bands from the UltraVISTA DR4 \citep{Mccracken2012}, and IRAC ch1 and ch2 data from the Spitzer Extended Deep Survey (SEDS; \citealt{Ashby2013,Ashby2018}).

Meanwhile, we employed 57 X-ray selected AGNs at $z_\mathrm{spec}=$ 1.5--2.5 in the GOODS-S field ($\sim0.05$ deg$^2$) down to $6.4\times10^{-18}$ erg~s$^{-1}$cm$^{-2}$ in the soft band and $2.7\times10^{-17}$ erg~s$^{-1}$cm$^{-2}$ in the hard band \citep{Luo2017,Liu2017}, which are significantly deeper than those in the Spiderweb protocluster (see also figure~4 in \citealt{Tozzi2022a}).
It should be noted here that we used their unabsorbed X-ray luminosity only in the hard band and photon indexes for the SED fitting due to the data availability in \citet{Liu2017}.
Photometric data were taken from the 3D-HST catalogue \citep{Skelton2014}, which is a collection of 
VLT/VIMOS $UR$ bands from \citet{Nonino2009}, $UBVR_cI$ bands from the Garching-Bonn Deep Survey \citep{Erben2005,Hildebrandt2006}, median bands from MUSYC survey \citep{Cardamone2010}, HST/ACS F435W,F606W band data from GOODS \citep{Giavalisco2004}, F606W,F814W,F125W,F160W bands from CANDELS \citep{Grogin2011,Koekemoer2011}, F140W band from 3D-HST \citep{Brammer2012}, $JHK_s$ bands from GOODS and FIREWORKS \citep{Wuyts2008,Retzlaff2010}, $JK_s$ bands from TENIS \citep{Hsieh2012}, and IRAC ch1 and ch2 data from SEDS \citep{Ashby2013}.
We then performed the SED fitting in the exact same manner as for the HAE sample (see Section~\ref{s22}).
For a relatively fair comparison, we employed the band photometry in the X-ray and the $U$-to-4.5 $\mu$m bands for these field samples.

% figure 5
\begin{figure}
\centering
\includegraphics[width=0.97\columnwidth]{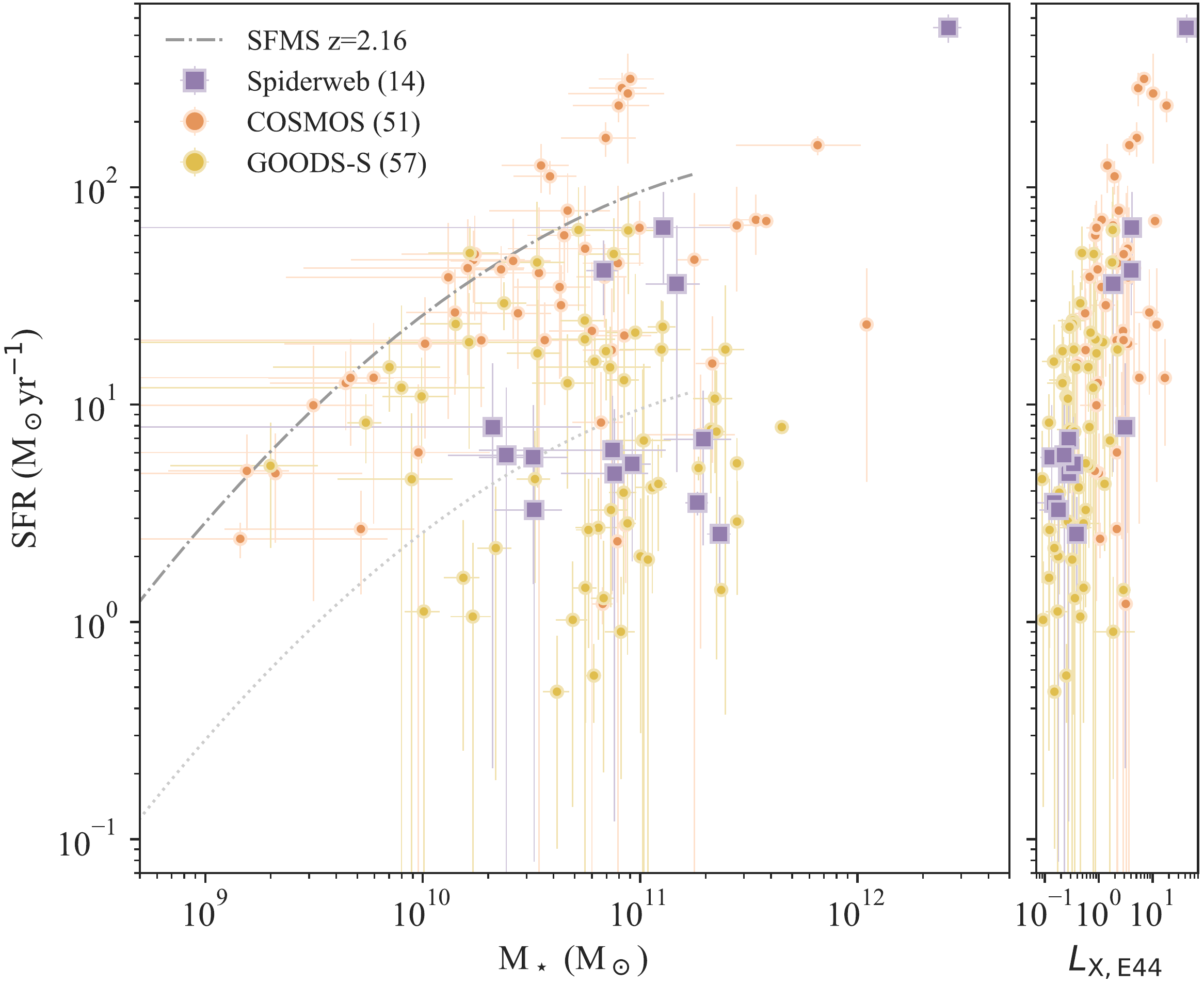}
\caption{
The left panel is the same as Figure~\ref{fig3} but for 14 X-ray HAEs in the Spiderweb protocluster (purple squares) and 51 and 57 X-ray AGNs at $z=$ 1.5--2.5 in the COSMOS and GODDS-S fieldsm respectively (orange and yellow circles).
The right panel shows SFRs versus X-ray luminosity in the hard band (2--10 keV) in the unit of $1\times10^{44}$ erg~s$^{-1}$ for the same samples as in the left panel.
}
\label{fig4}
\end{figure}

Figure~\ref{fig4} shows the resultant comparison plot between the Spiderweb protocluster and the COSMOS and GOODS-S fields on the SFR--M$_\star$ plane.
Despite the use of similar data and the SED modelling, the figure indicates a clear difference. 
Specifically, X-ray HAEs in the Spiderweb protocluster are preferentially located below the star-forming main sequence compared to those in the COSMOS field at $z=$ 1.5--2.5, while they show a better agreement with the GOODS-S samples at $z=$ 1.5--2.5.
The lack of X-ray AGNs with low specific SFRs (sSFRs) in the COSMOS field is likely due to the sampling bias, such as a different X-ray flux limit and a different spectroscopic identification method, rather than the environmental dependence.
X-ray AGNs with low sSFRs in the Spiderweb protocluster tend to have lower X-ray luminosity just below the flux limit in the COSMOS field (Fig.~\ref{fig4}).
\citet{Ito2022} have detected faint AGN emissions in the composite X-ray image of quiescent galaxies in the same COSMOS field at a similar redshift, which further supports this scenario.
Their composite X-ray flux in the soft band is $\sim3\times10^{-17}$ erg~s$^{-1}$cm$^{-2}$ in passive galaxies with M$_\star=1\times10^{11}$ M$_\odot$ at $z=$ 2.0--2.5 (via private communication with \citealt{Ito2022}).
It is thus reasonable to consider that we are simply missing such passively evolving galaxies with faint X-ray emissions in the COSMOS field.
In addition, the sampling bias of redshift identifications may affect the comparison analysis.
While the X-ray AGNs in the protocluster have been selected based on H$\alpha$ + [N{\sc ii}] lines with near-infrared observations, those in the COSMOS field have been originally selected based on the optical spectroscopy with Keck/DEIMOS \citep{Hasinger2018}.
This difference in sample selection may lead the apparent difference between the Spiderweb protocluster and the COSMOS field, as shown in Figure~\ref{fig4}.

Indeed, we also observed quiescent hosts with low-luminosity AGNs in the GOODS-S field, where the X-ray flux limit is much deeper than those in the COSMOS field and the Spiderweb protocluster (Fig.~\ref{fig4}).
However, we should note that there is no tight correlation between SFRs and X-ray luminosity within the range of $<1\times10^{44}$ erg~s$^{-1}$, suggesting that instantaneous AGN activity is not connected with star formation quenching in massive galaxies (see also discussion in \citealt{Terrazas2020,Piotrowska2022,Ward2022,Bluck2023a,Bluck2023b}). 
Taken together, we conclude that the discovery of passive HAEs with low-luminosity AGNs in the Spiderweb protocluster would not be caused environmental effects. 
Such objects can be found even in the general field as long as one has very deep X-ray data, and rather, they may be more abundant than we previously thought.
On the other hand, the Spiderweb protocluster may yet have a high fraction of such passive HAEs with low-luminosity AGNs since high-density environments can provoke galaxy mergers hence promote the mass growth of black holes \citep{Okamoto2000,Gottlober2001,Hine2016}. 
To this end, a homogeneous and deep observation in both H$\alpha$ and X-ray will be required for deriving the abundance of star-forming and passive galaxies with X-ray AGNs down to the low X-ray luminosity in a quantitative manner.

\subsection{Impacts of AGNs on H-Alpha emission}\label{s42}

We have argued so far that 9 out of 14 X-ray HAEs seem to be post star-forming galaxies despite the presence of H$\alpha$ (+ [N{\sc ii}]) line emissions ($>4\times10^{-17}$ erg~s$^{-1}$cm$^{-2}$, \citealt{Shimakawa2018}).
In this subsection, we discuss how large AGN contributions make up for their H$\alpha$ fluxes detected by the narrow-band imaging.

\begin{figure}
\centering
\includegraphics[width=0.85\columnwidth]{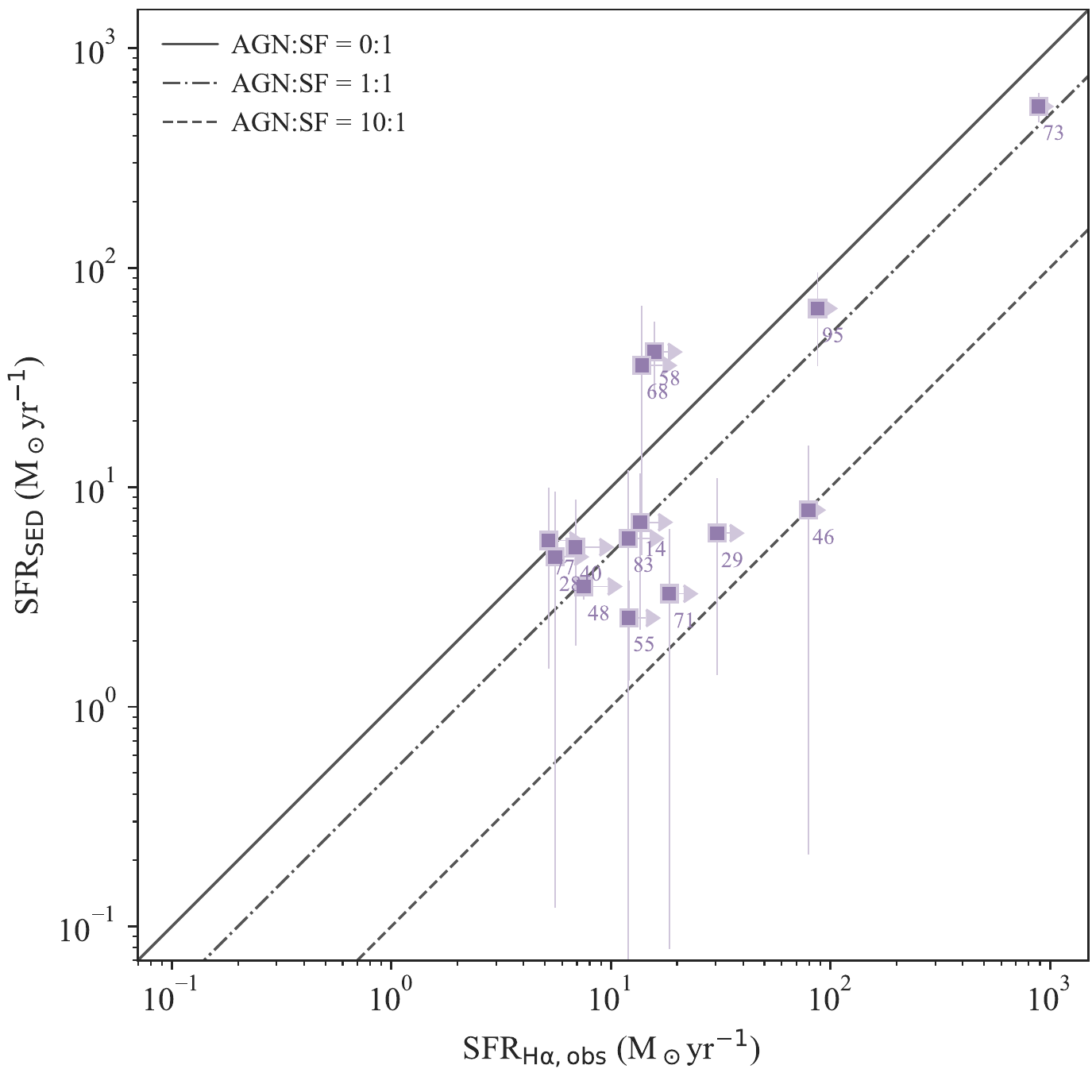}
\caption{
SED-based SFRs (SFR$_\mathrm{SED}$) versus H$\alpha$-based SFRs without dust correction (SFR$_\mathrm{H\alpha,obs}$) of 14 X-ray HAEs.
Their SFR$_\mathrm{H\alpha}$ values are simply converted through the \citet{Kennicutt1998} prescription based on observed narrow-band fluxes, assuming 30\% contributions of [N{\sc ii}] lines.
The solid, dot-dash, and dashed lines respectively assume 100, 50, and 10\% contributions of star formation to their observed H$\alpha$ fluxes, although the assumed fractions should be minimum, given dust attenuation.
}
\label{fig5}
\end{figure}

Comparing two independently-obtained SFRs from the SED fitting and observed H$\alpha$ fluxes is a convenient way to examine the AGN contribution (Fig.~\ref{fig5}).
In the latter case we estimated H$\alpha$-based SFRs (SFR$_\mathrm{H\alpha,obs}$) using the \citet{Kennicutt1998} prescription with additional scaling factor of 0.59 to convert to the \citet{Chabrier2003} IMF.
We adopted H$\alpha$ fluxes of X-ray HAEs from their narrow-band fluxes obtained by \citet{Shimakawa2018}, assuming 30\% of [N{\sc ii}] flux contribution to the narrow-band fluxes.
It should be noted that we do not take dust extinction into account for the sake of simplicity, that is, the H$\alpha$-based SFR assumed here should be minimum.
For reference, we expect that typical H$\alpha$ extinctions of these massive galaxies with M$_\star\sim1\times10^{11}$ M$_\odot$ would be $\sim2$ mag if they were normal star-forming galaxies \citep{Shimakawa2015}, which increase their SFRs by a factor of 6.31 when applying dust correction.
Figure~\ref{fig5} shows that a majority of the X-ray HAEs require AGN contributions to their H$\alpha$ + [N{\sc ii}] lines for making them consistent with their SED-based SFRs. 
Particularly, at least one of those should be affected by approximately 90\% or more even if dust correction is not taken into account. 
We thus conclude that the bulk of H$\alpha$ emission detected in the previous narrow-band imaging \citep{Shimakawa2018} would originate from AGNs and not host galaxies.
A follow-up deep near-infrared spectroscopy will be helpful to further constrain their emission line contributions as discussed in Section~\ref{s43}.
The forthcoming Pa$\beta$ imaging with JWST/NIRCam \citep{Dannerbauer2021} could be also able to spatially decompose emission-line contributions from AGNs and host galaxies.

% figure 6

\begin{figure}
\centering
\includegraphics[width=0.85\columnwidth]{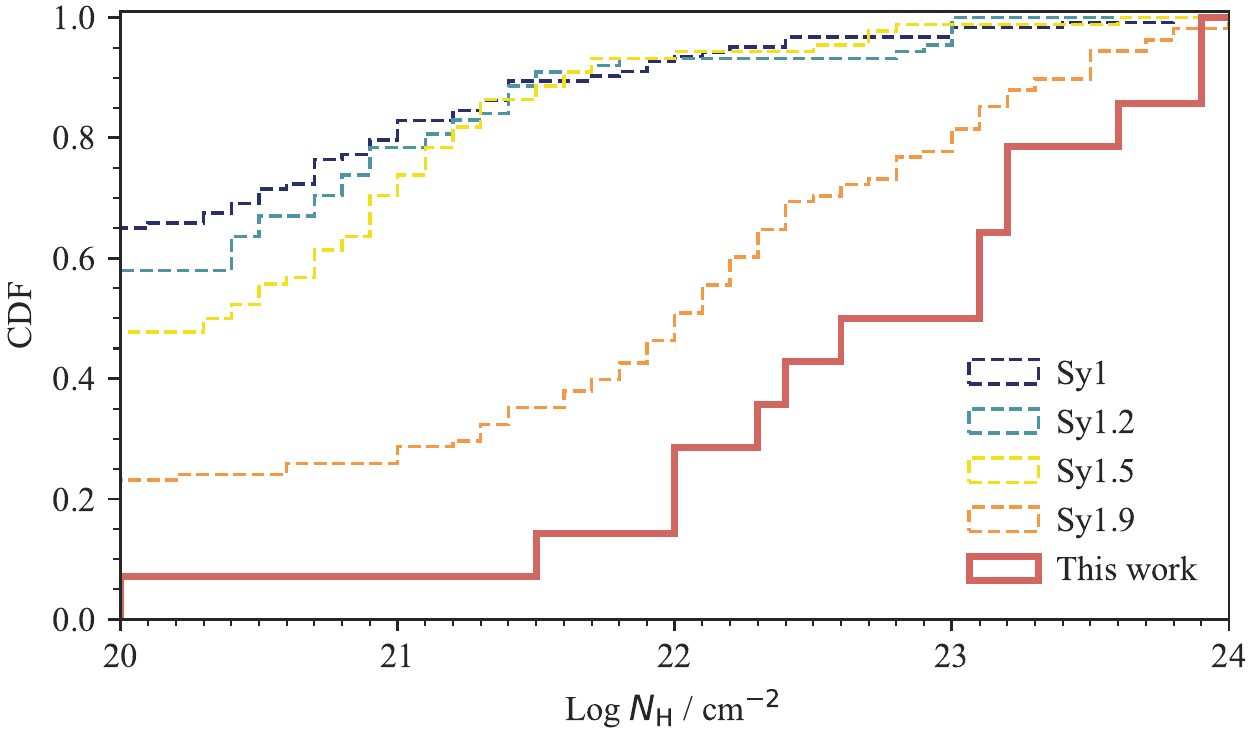}
\caption{
The red thick line shows the normalised cumulative distribution function of H{\sc i} column densities for the 14 X-ray HAEs from \citet{Tozzi2022a}.
We here assumed $N_\mathrm{H}=10^{24}$ cm$^{-2}$ for two Compton-thick AGNs and $N_\mathrm{H}=10^{20}$ cm$^{-2}$ for 1 AGN with $N_\mathrm{H}=0$ cm$^{-2}$ in the literature for the sake of visibility.
We also adopted the upper limits ($N_\mathrm{H}\sim3\times10^{22}$ cm$^{-2}$) for 3 AGNs (see \citealt[table~5]{Tozzi2022a}). 
Additionally, for reference, the column densities of various types of Seyfert AGNs at $z<0.4$ are shown by dashed lines \citep{Mejia-Restrepo2022}, where Seyfert 1, 1.2, 1.5, and 1.9 sources are coloured blue, cyan, yellow, and orange, respectively.
}
\label{fig6}
\end{figure}

Dust attenuation reduces observed H$\alpha$ fluxes from not only star-forming regions but also AGNs. 
\citet{Tozzi2022a} reported that X-ray HAEs in the Spiderweb protocluster tend to have high H{\sc i} column densities, which suggests that their H$\alpha$ fluxes (and perhaps FWHM) would be significantly suppressed as observed in type-1.9 AGNs at lower redshifts \citep{Mejia-Restrepo2022,Ricci2022}.
Figure~\ref{fig6} presents normalised cumulative distribution functions (CDFs) of the H{\sc i} column densities for the X-ray HAEs and X-ray selected AGNs at $z<0.4$ from the BAT AGN Spectroscopic Survey (BASS/DR2).
We have collected the column density values of the X-ray HAEs from \citet[table~5]{Tozzi2022a} and those at low redshifts from a compilation by \citet{Ricci2017a,Oh2018,Mejia-Restrepo2022}.
Although we have to treat comparison results in the figure with caution, given the potential sampling bias, H{\sc i} column densities are greater than $1\times10^{22}$ cm$^{-2}$ for a majority of the X-ray HAEs, which is more similar to those of Seyfert 1.9 AGNs than those of Seyfert $<1.9$.
This suggests that familiar broad line emissions such as Mg{\sc ii} and H$\beta$ would not be visible in the X-ray HAE sample, and even H$\alpha$ broad lines may be suppressed.
Such a trend also agrees with the recent findings that high-redshift AGNs tend to be dust obscured by their massive host galaxies \citep{Gilli2022,Silverman2023}.

% figure 7

\begin{figure}
\centering
\includegraphics[width=0.85\columnwidth]{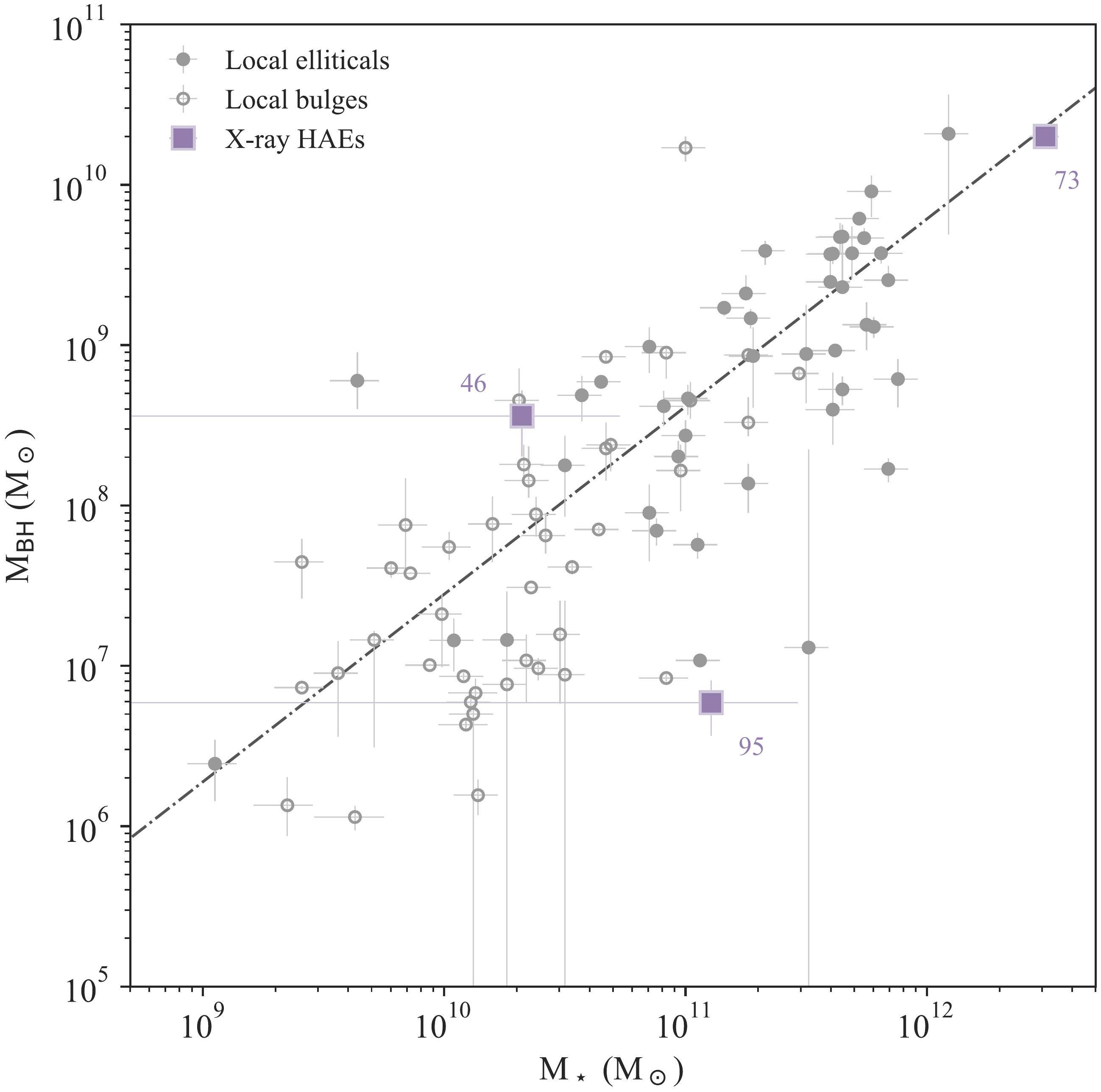}
\caption{
Black hole mass versus stellar mass for local elliptical galaxies (filled circles) and classical and pseudo-bulges (opened circles) from \citet{Kormendy2013}.
We here plot the tentative black hole mass measurements for three X-ray HAEs (ID=46, 73, 95; Table~\ref{tab1}) as shown by the magenta squares.
Black hole masses of ID=46 and 95 are newly obtained through the \citet{Mejia-Restrepo2022} prescription, while we adopted the estimation by \citet{Tozzi2022a} for the Spiderweb radio galaxy (ID=73).
The dot-dash line depicts the local M$_\mathrm{BH}$--M$_\star$ relation \citep{Kormendy2013} scaled by \citet{Reines2015}.  
}
\label{fig7}
\end{figure}

In fact, \citet{Perez-Martinez2023} obtained H$\alpha$ spectra for 8 out 14 X-ray HAEs with the VLT K-band Multi Object Spectrograph (KMOS, \citealt{Sharples2013}), including ID=14, 28, 29, 46, 55, 58, 71, and 95 on Table~\ref{tab1}, but only two of them (ID=46, 95) show relatively broad components with $>700$ km~s$^{-1}$.
We tentatively estimated their black hole masses (M$_\mathrm{BH}$) based on the best-fit H$\alpha$ broad line properties from \citet[figure~3]{Perez-Martinez2023}, through the prescription in \citet[equation~2]{Mejia-Restrepo2022}, which is modified from the \citet{Greene2005} calibration,
\begin{equation}
\mathrm{M_{BH}} = 10^{6.43}\times\left(\frac{L_\mathrm{H\alpha}}{10^{42}~\mathrm{erg~s^{-1}}}\right)^{0.55}\left(\frac{\mathrm{FWHM_{H\alpha}}}{10^{3}~\mathrm{km~s^{-1}}}\right)^{2.06}\mathrm{M_\odot}.
\label{eq1}
\end{equation}
Obtained black hole masses are M$_\mathrm{BH}=3.6\pm1.6\times10^8$ and $5.9\pm2.2\times10^6$ M$_\odot$ in ID=46 and 95, respectively (Fig.~\ref{fig7}).
Both AGNs approximately follow the M$_\mathrm{BH}$--M$_\star$ relation of local elliptical galaxies and classical and pseudo-bulges \citep{Kormendy2013,Reines2015} within the margin of error.
As discussed above, these tentative measurements may underestimate their black hole masses because of the dust attenuation.
We also confirmed that the Spiderweb galaxy (ID=73) tracks the local M$_\mathrm{BH}$--M$_\star$ relation (Fig.~\ref{fig7}) when we adopted M$_\mathrm{BH}=2\times10^{10}$ M$_\odot$ reported by \citet{Tozzi2022a}.

% figure 8

The current dust-reddening issue prevents us from investigating their black hole masses as noted by \citet{Mejia-Restrepo2022,Ricci2022}.
The blending issue with the [N{\sc ii}] doublet and the limited signal-to-noise ratio are additional issues for accurately determining the broad line components.
Under such circumstances, for instance, a spectroscopic observation of Pa$\beta$ at $\lambda_\mathrm{obs}=4.05$ $\mu$m broad lines with the JWST Near Infrared Spectrograph (NIRSpec) will be a powerful solution to reliably constrain their dust-obscured black hole masses, where the more luminous Pa$\alpha$ line at $\lambda_\mathrm{obs}=5.93$ $\mu$m is out of the wavelength coverage of JWST/NIRSpec.
In addition, the forthcoming Pa$\beta$ narrow-band imaging with the JWST Near Infrared Camera (NIRCam, \citealt{Rieke2005}) on the Spiderweb protocluster may help us spatially resolve dust-obscured line emissions from AGNs and star formation.
These prospective follow-up analyses can provide useful insights into the growth history of super-massive black holes within red sequence galaxies in the present-day galaxy clusters.

\subsection{Spectroscopic characterisation}\label{s43}

Fortunately, we obtained the deep near-infrared spectral data from Keck/MOSFIRE for one of the X-ray HAEs, ID=55 (hereafter X-HAE-55), with M$_\star=2.32\pm0.29\times10^{11}$ M$_\odot$ and SFR $=2.54\pm1.23$ M$_\odot$yr$^{-1}$ (Table~\ref{tab1}).
The X-HAE-55 was previously discussed as one of the extremely red objects and hence a quiescent galaxy in the Spiderweb protocluster by \citet{Doherty2010,Tanaka2013}.
Interestingly, H$\alpha+$[N{\sc ii}] emission with the flux of $1\times10^{-16}$ erg~s$^{-1}$cm$^{-2}$ has been detected by the narrow-band imaging \citep{Koyama2013a,Shimakawa2018}, which is confirmed by the recent work with the VLT/KMOS spectroscopy \citep{Perez-Martinez2023}.
This work indicated that the bulk of H$\alpha$ + [N{\sc ii}] emission should come from the AGN based on the AGN$+$galaxy SED fitting (see also discussion in \citealt{Tozzi2022a}). 
Such a passive galaxy with AGN emission has also been reported at $z=4.6$ \citep{Carnall2023}.
This last subsection provides a complementary analysis on the X-HAE-55 based on the deep spectral data.

\begin{figure}
\centering
\includegraphics[width=0.95\columnwidth]{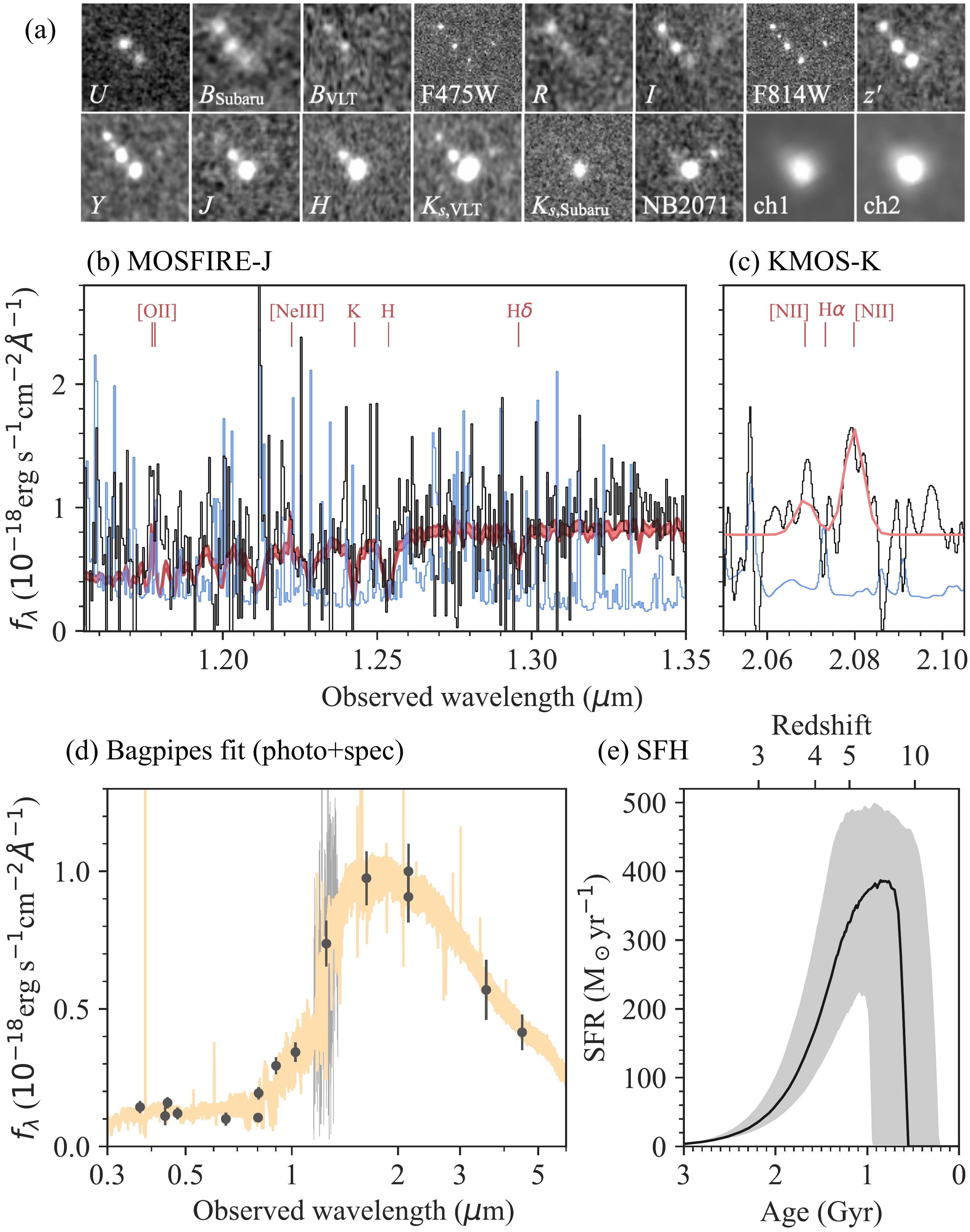}
\caption{
(a) Cutout images of the X-HAE-55 in the $U$-to-ch2 bands.
(b) The black and blue lines indicate the observed spectrum and noise in the MOSFIRE $J$-band, respectively, where the spectrum is binned with 4 spectral pixels.
The red filled region depicts the 68 percentile distribution from the {\tt bagpipes} SED fitting.
(c) Same as in (b) but for the KMOS $K$-band data.
The red curve is the best-fit [N{\sc ii}]$\lambda\lambda$6550,6585 doublet, assuming the fixed flux ratio of 1:3.
The spectrum is convolved by a Gaussian filter with $\sigma=2$ spectral pixels from the original spectrum in \citet[figure~3]{Perez-Martinez2023}.
(d) The orange region indicates the 68 percentile distribution from the best-fit SED with {\tt Bagpipes} \citep{Carnall2018,Carnall2019}, where photometric fluxes and $J$-band spectrum are shown by the black circles with error-bars and the grey line, respectively. 
(e) The best-fit SFH (black line) and the 68 percentile uncertainty (grey areas) of the X-HAE-55 as a function of the cosmic age and redshift.
}
\label{fig8}
\end{figure}

The spectrum of HAE ID=55 was obtained in 2015 January 26--27 using Keck/MOSFIRE \citep{McLean2010,McLean2012} through the Subaru-Keck time exchange framework. 
The primary targets were quiescent galaxy candidates and the $J$-band grating was used to cover the wavelength range of $\lambda=$ 1.17--1.35 $\mu$m, which includes the rest-frame 4000\AA\ break, with a spectral resolution of $R\sim3300$.
The observing conditions were good with a typical seeing size of about 0.7 arcsec in the $J$-band.
The standard ABBA nodding was applied along the slit with an exposure time of 120 seconds at each position.
The data were reduced in the standard manner using the MOSFIRE Data Reduction Pipeline ({\tt MOSFIRE-DRP}; see \citealt{Steidel2014} for details). 
The flux was calibrated against A-type stars observed during the nights.
The total integration time was 3 hours. 
In this work, we apply 4-pixel binning along the wavelength to increase the signal-to-noise ratio, and the spectrum is shown in Figure~\ref{fig8}.

We then performed the SED fitting with the {\tt Bagpipes} code (\citealt{Carnall2018,Carnall2019}, version~1.0.2) based on the combined dataset of photometry from the $U$ to ch1 bands (Section~\ref{s22}) and the reduced $J$-band spectrum.
The {\tt Bagpipes} code relies on the Python module {\tt MultiNest} for Bayesian analysis \citep{Buchner2014,Feroz2019}. 
We rescaled band photometry to exclude the AGN contribution based on flux ratios of the best-fit galaxy SED to the AGN component obtained from {\tt X-CIGALE} (Fig.~\ref{fig2}).
The redshift was fixed within the range of $z=$ 2.150--2.165 given Ca{\sc ii} H and K absorption lines and tentative detections of the [O{\sc ii}] doublet (Fig.~\ref{fig8}). 
It should be noted that this redshift range falls out of the previous measurement in the literature ($z=2.1694$, \citealt{Perez-Martinez2023}), derived from an apparent H$\alpha$ line in the $K$-band spectrum with the VLT/KMOS spectroscopy.
However, it is turned out that the referenced line should be [N{\sc ii}]$\lambda6585$ line not H$\alpha$ line by comparing with the MOSFIRE $J$-band spectrum (Fig.~\ref{fig8}).
The H$\alpha$ line emission cannot be seen in the current KMOS data, meaning that the host galaxy is quenched and even AGN-induced H$\alpha$ line may be hidden by stellar absorption and/or sky noise.
In fact, the obtained [N{\sc ii}] fluxes account only for 62\% of the narrow-band flux of $1\times10^{-16}$ erg~s$^{-1}$cm$^{-2}$ taken from \citet{Shimakawa2018}, and thus, we should be missing unignorable line contributions from the host galaxy and AGN due to the limited signal-to-noise ratio.
Despite the missing components, the line spectrum resembles Low Ionization Nuclear Emission-line Regions (LINERs; \citealt{Heckman1980}) given a high [N{\sc ii}]/H$\alpha$ ratio ($\gtrsim1$). 
Such a LINER-like feature in post star-forming galaxies with low-luminosity AGN is compatible with the recent AGN quenching scenario suggested from the modern hydrodynamic cosmological simulations \citep{Terrazas2020,Piotrowska2022,Bluck2023a,Bluck2023b}, where cumulative energy injections from kinetic AGN feedback at low Eddington ratios play a pivotal role in star formation quenching.
Considering all the factors, the other passive HAEs could also be LINERs, although further investigations must be needed.

The derived best-fit SED and SFH are summarised in Figure~\ref{fig8}.
Because of the similar parameter setting in the SED modelling, obtained stellar mass $=2.27\pm0.25\times10^{11}$ M$_\odot$ and SFR $=4.97\pm1.95$ M$_\odot$yr$^{-1}$ are consistent with those from {\tt X-CIGALE} (Table~\ref{tab1}).
In addition, the best-fit SED spectrum well reproduces stellar absorption lines of Ca{\sc ii} H\&K and H$\delta$ seen in the MOSFIRE $J$-band spectrum.
Moreover, the best-fit SFH with associated errors suggest that the X-HAE-55 was intensively formed during $z=$ 4--10 (or age of $\sim2.0\pm0.5$ Gyr) like high-$z$ submillimetre galaxies (e.g., \citealt{Smail1997,daCunha2015,Valentino2020}), and then shifted to the quiescent phase over the last $\sim1$ Gyr before the observed redshift ($z=2.16$).
This formation history is consistent with those reported in previous studies for extremely red objects in the Spiderweb protocluster \citep{Kodama2007,Doherty2010,Tanaka2013}.
Because we see a clear AGN sign in this system, AGN feedback is a favourite scenario to explain its star formation quenching (e.g., \citealt{Alexander2012,Fabian2012,Harrison2017} and references therein, but see \citealt{Terrazas2020,Piotrowska2022,Bluck2023a,Bluck2023b}).
However, we still lack sufficient proof to conclude that with the current data, which needs to be addressed in the future.

% figure 9

%%%%%%%%%%%%%%%%% CONCLUSIONS %%%%%%%%%%%%%%%%%%
\section{Conclusions}\label{s5}

Taking the recent deep Chandra X-ray observation \citep{Tozzi2022a} as its basis and using rich datasets compiled from the various previous studies across the multi-wavelength, this work revisited the stellar mass and SFR measurements for 14 AGN-host HAEs in the Spiderwerb protocluster at $z=2.16$ \citep{Shimakawa2018}. 
Our result of the SED fitting indicates that about half of massive HAEs with M$_\star>2\times10^{10}$ M$_\odot$ host X-ray AGNs (14/30).
Further, nine massive HAEs with X-ray counterparts and hence X-ray AGNs tend to show low SFRs significantly below ($\sim1$ dex) the star-forming main sequence traced by the other HAEs. 
This suggests that the bulk of their H$\alpha$ (+ [N{\sc ii}]) emissions would originate from AGNs rather than from star formation of host galaxies.
Here, the SED fitting algorithm {\tt X-CIGALE} with the X-ray module \citep{Boquien2019,Yang2020} and the deep photometry in the X-ray bands \citep{Tozzi2022a} played the critical role of reproducing such an intriguing trend.
Although a follow-up investigation is still needed to directly confirm their quiescence, our follow-up Keck/MOSFIRE and VLT/KMOS data spectroscopically supports the quiescent nature for at least one of these AGN hosts.

In the historical context, the Spiderweb protocluster was first confirmed by overdensities of Ly$\alpha$ and H$\alpha$ (+ [N{\sc ii}]) emitters \citep{Kurk2000,Pentericci2000,Kurk2004a,Kurk2004b}. 
Follow-up (spectro-)photometric studies then discovered an apparent red sequence formed by extremely red objects with low SFRs \citep{Kodama2007,Doherty2010,Tanaka2010,Tanaka2013}. 
Further intensive surveys found H$\alpha$ (+ [N{\sc ii}]) and dust emissions from most of these red objects. 
This led to the understanding that the red galaxies in the protocluster would be dominated by dusty starbursts that are on the verge of becoming bright red sequence galaxies \citep{Koyama2013a,Koyama2013b,Dannerbauer2014,Shimakawa2018}. 
The results of this study suggest that significant fractions of line emissions in the red massive HAEs actually originate from AGNs not star formation, thereby providing a reasonable agreement on such observational results reported thus far.

The results indicate that a third of massive HAEs with M$_\star>2\times10^{10}$ M$_\odot$ are quiescent galaxies with low-luminosity AGNs (9/30). 
This implies that AGNs may significantly involve passive galaxies in the Spdierwerb protocluster as only three additional quiescent members with spectroscopic redshifts have been reported so far \citep{Tanaka2013}.
Such a startling evidence of ubiquitous AGNs in the red sequence in the Spiderweb protocluster suggests that AGNs might be physical drivers of their star formation quenching. 
Further constraints on star formation and AGN activity, as well as physical interplay between host galaxies and AGNs in the protocluster, remains to be addressed in future studies. 
Some of these may be partially addressed by the forthcoming Pa$\beta$ imaging with JWST/NIRCam \citep{Dannerbauer2021} and/or further follow-up observations with such as NIRSpec and MIRI on JWST.

%%%%%%%%%%%%%%%%% Acknowledgements %%%%%%%%%%%%%%%%%%

\section*{Acknowledgements}

We thank anonymous referee for useful comments.
The data are collected at the Subaru Telescope operated by the NAOJ, the Hubble Legacy Archive, which is a collaboration between the STScI/NASA, the ST-ECF/ESA and the CADC/NRC/CSA, and the ESO Science Archive Facility under programme ID 383.A-0891.
In addition, this research has made use of the NASA/IPAC Infrared Science Archive, which is funded by the National Aeronautics and Space Administration and operated by the California Institute of Technology.
This work is in part based on observations taken by the 3D-HST Treasury Program (GO 12177 and 12328) with the NASA/ESA HST, which is operated by the Association of Universities for Research in Astronomy, Inc., under NASA contract NAS5-26555. 
Also. this work is in part based on observations collected at the European Southern Observatory under ESO programme ID 179.A-2005 and on data products produced by CALET and the Cambridge Astronomy Survey Unit on behalf of the UltraVISTA consortium.
We are honoured and grateful for the opportunity of observing the universe from Maunakea, which has the cultural, historical, and natural significance in Hawaii.

We would like to thank Editage (\url{www.editage.com}) for English language editing.
This work is supported by a Waseda University Grant for Special Research Projects (2023C-590) and MEXT/JSPS KAKENHI Grant Numbers (23H01219 and 18H03717).
Numerical computations were in part carried out on Cray XC50 at Center for Computational Astrophysics, National Astronomical Observatory of Japan.
This work made extensive use of the following tools, {\tt NumPy} \citep{Harris2020}, {\tt Matplotlib} \citep{Hunter2007}, {\tt TOPCAT} \citep{Taylor2005}, {\tt Astopy} \citep{AstropyCollaboration2013,AstropyCollaboration2018}, and {\tt pandas} \citep{Reback2022}.

%%%%%%%%%%%%%%%%%%%%%%%%%%%%%%%%%%%%%%%%%%%%%%%%%%
\section*{Data Availability}

This work is based on compilation data of previous studies (Table~\ref{tab2}, see also \citealt{Shimakawa2017,Shimakawa2018}).
Each science-ready data will be shared on reasonable request to the corresponding author.
Moreover, all the data used in this work can be obtained through the Subaru Mitaka Okayama Kiso Archive (SMOKA) system, the Keck Observatory Archive, the Hubble Legacy Archive, the NASA/IPAC Infrared Science Archive, and the ESO Science Archive Facility.



\bibliographystyle{mnras}
\bibliography{rs23a} % if your bibtex file is called example.bib

%\begin{thebibliography}{99}
%\end{thebibliography}

%%%%%%%%%%%%%%%%%%%%%%%%%%%%%%%%%%%%%%%%%%%%%%%%%%

%%%%%%%%%%%%%%%%% APPENDICES %%%%%%%%%%%%%%%%%%%%%

\appendix

\section{Comparison of our measurements with those in previous work}\label{a1}

Given that our choices of stellar populations in the SFH were different from those in previous work, there were systematic differences in the derivation of physical properties such as stellar masses and SFRs.
We then discuss the consistency between SED-based SFRs and H$\alpha$-based SFRs of non-X-ray HAEs, and also systematic errors between previous stellar mass measurements of the entire HAEs in \citet{Shimakawa2018} and those in this work.

\begin{figure}
\centering
\includegraphics[width=0.9\columnwidth]{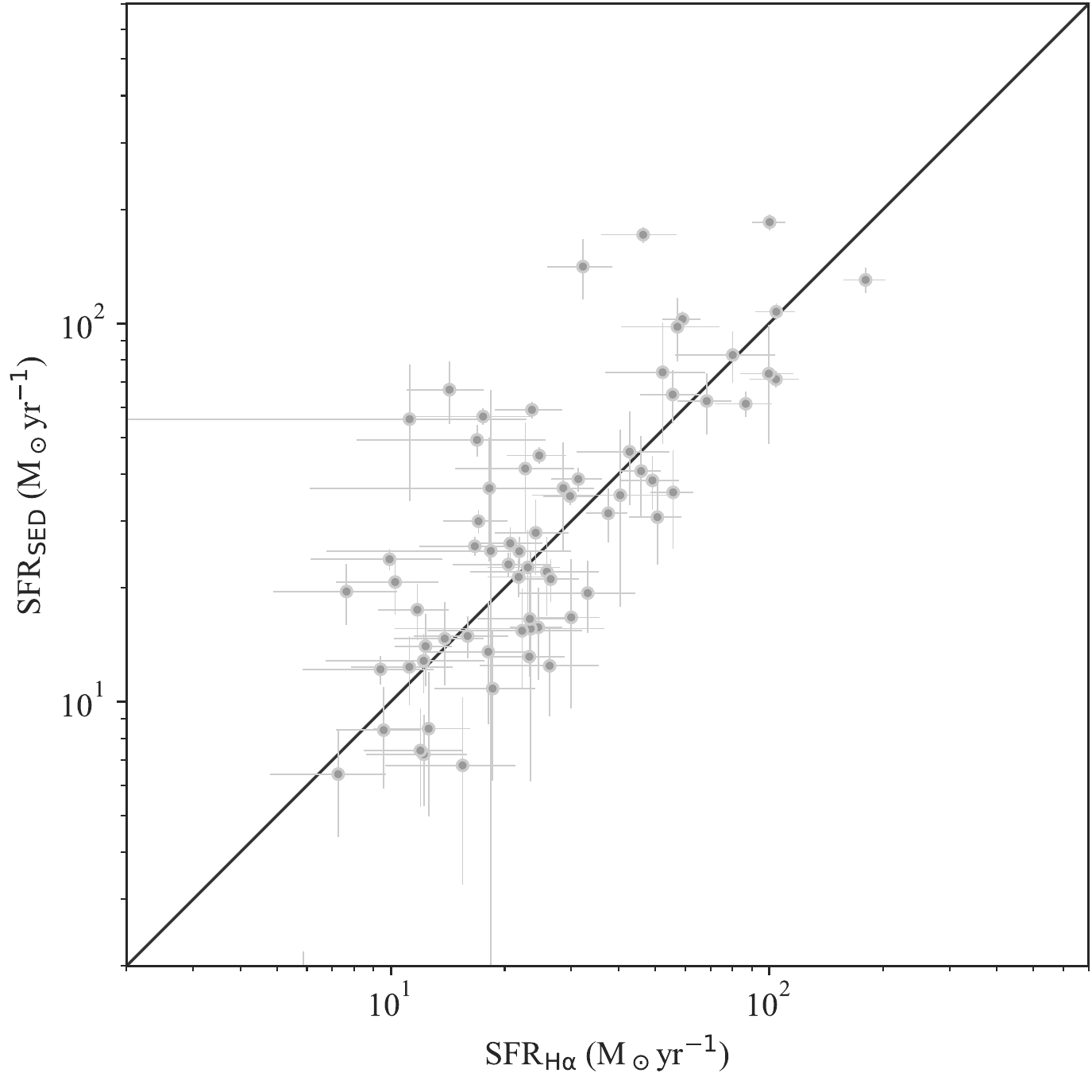}
\caption{
Comparison of SED-based SFRs with H$\alpha$-based SFRs for 70 non X-ray HAEs (grey circles with $1\sigma$ error-bars) in the Spiderweb protocluster at $z=2.16$.
}
\label{fig1a}
\end{figure}

Figure~\ref{fig1a} shows the comparison between SED-based SFRs and H$\alpha$-based SFRs for 70 ($=84-14$) HAEs \citep{Shimakawa2018} without the X-ray counterparts \citep{Tozzi2022a}.
While their SED-based SFRs are the same as shown in Figure~\ref{fig2}, H$\alpha$-based SFRs are derived from narrow-band fluxes in \citet{Shimakawa2018}, assuming 30\% of flux contributions from the [N{\sc ii}] doublet and dust extinctions from the {\tt X-CIGALE} SED fitting.
We confirmed that, given the adopted SED modelling, these two different estimates are broadly consistent with each other, although there are deviations from the 1:1 relation.

% figure 1a

We then cross-checked our stellar mass measurements with those in the previous work \citep{Shimakawa2018}, which derived the stellar mass based on the {\tt FAST} SED fitting code \citep{Kriek2009}.
The major change in the model assumption compared to previous work is the exclusion of young stellar populations with age $\lesssim100$ Myr in the main population; this is because this work set up the model parameters following those in \citet{Pearson2017}.
The reason for removing the young populations is to derive the reasonable SED-based SFRs for the HAE samples; otherwise, we obtain very high SED-based SFRs that do not match their H$\alpha$-based SFRs and the star-forming main sequence at $z=2.16$ (e.g., \citealt{Tomczak2016}).
This deviation is attributed to the fact that the SED spectra are less sensitive to older stellar populations in general. 
We should note that such a different model assumption leads to systematic deviations in stellar mass and SFR measurements but does not affect our conclusion as the obtained dependencies of star formation on AGN activities (Fig.~\ref{fig3} and \ref{fig4}) do not change.

\begin{figure}
\centering
\includegraphics[width=0.9\columnwidth]{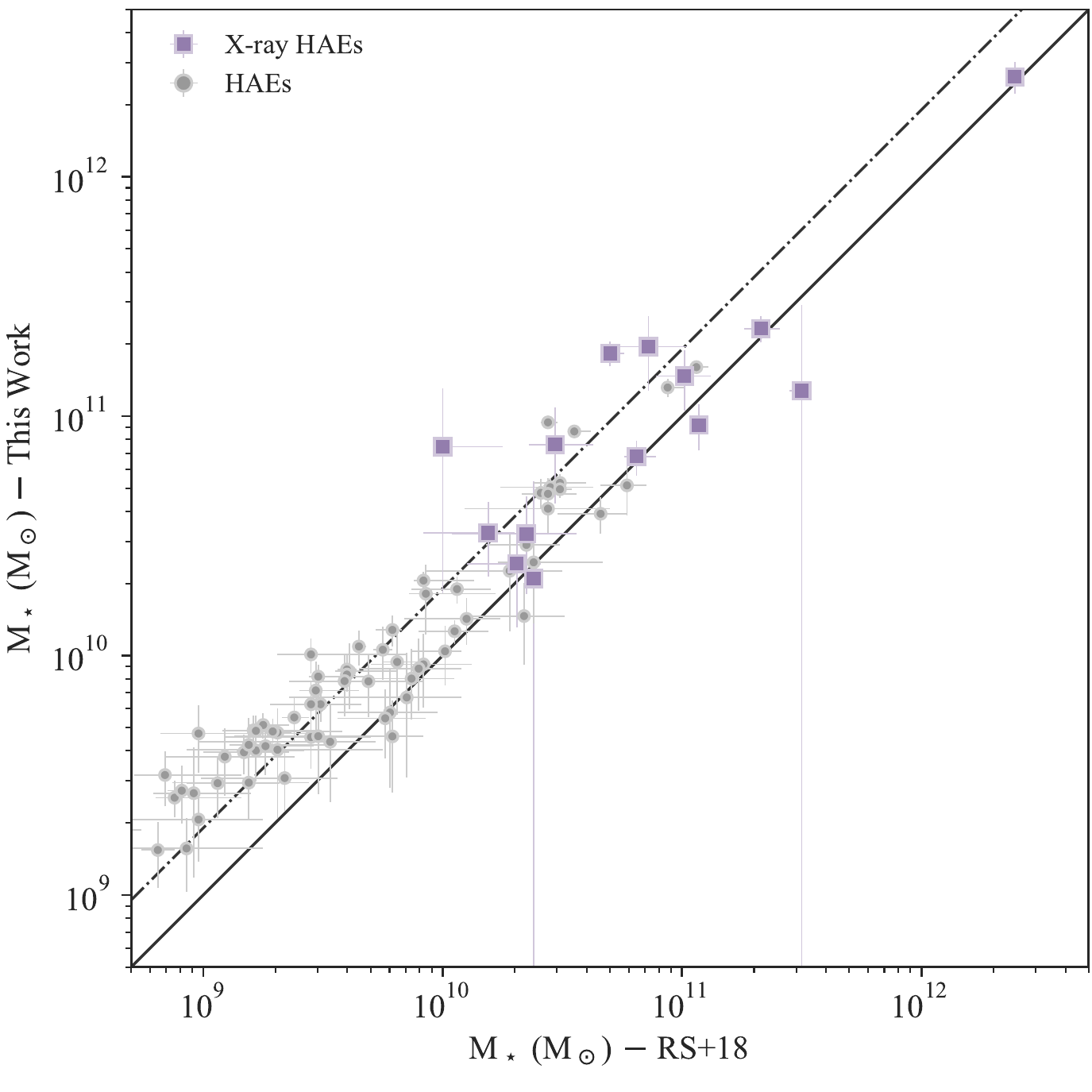}
\caption{
Comparison of the stellar mass estimation between this work and \citet{Shimakawa2018} for the entire 84 HAEs.
The black solid and dash-dot lines depict the 1:1 relation and that with an offset of $+0.28$ dex.
}
\label{fig2a}
\end{figure}

The extent of deviations in stellar mass measurements between this work and \citet{Shimakawa2018} can be found in Figure~\ref{fig2a}. 
This work tends to obtain higher stellar masses for HAEs with a typical offset value of 0.28 dex, which increases to 0.34 dex when we limit the sample to low-mass HAEs with M$_\star<1\times10^{10}$ M$_\odot$.
We also obtained significantly higher stellar masses in some X-ray HAEs by adding the AGN components in the SED fitting.
Meanwhile, this work would over-estimate stellar masses of some low-mass active HAEs with specific SFRs $\gtrsim10$ Gyr$^{-1}$, meaning their masses could be acquired in the last $\lesssim100$ Myr. This is because we do not consider young populations ($\leq100$ Myr) as the main stellar population in the SED modelling on the {\tt X-CIGALE} code \citep{Boquien2019,Yang2020}, though we include the late constant burst up to 20\% in the mass fraction as a compromise.
However, it should be noted that this potential issue does not affect the conclusion as these young low-mass HAEs are beyond the scope of this paper.

% figure 2a

\begin{comment}

% table 4
\begin{table*}
\centering
\caption{
Same as Table~\ref{tab1} but for all 84 HAEs updated by this work.
References are written in the following abbreviations (S24: this work, P23: \citealt{Perez-Martinez2023}, J21: \citealt{Jin2021}, S14: \citealt{Shimakawa2014}, C05: \citealt{Croft2005}, K04: \citealt{Kurk2004b}, K00: \citealt{Kurk2000}).
}
\label{tab4}
    \begin{tabular}{cccccccc} % four columns, alignment for each
    \hline
    ID & R.A. & Dec. & $z_\mathrm{spec}$ & Line & Reference & M$_\star$ [$10^{10}$ M$_\odot$] & SFR [M$_\star$yr$^{-1}$]\\
    \hline
    2  & 11:40:55.20 & $-$26:30:42.7 & 2.1447 & H$\alpha$  & P23 & $0.38\pm0.12$ & $15.71\pm4.30$\\
    4  & 11:40:59.81 & $-$26:30:42.6 & 2.1615 & H$\alpha$  & P23 & $1.09\pm0.18$ & $66.79\pm12.60$\\
    5  & 11:40:57.80 & $-$26:30:48.1 & 2.1635 & H$\alpha$  & P23 & $13.19\pm1.14$ & $107.3\pm5.4$\\
    ...\\
    \hline
\end{tabular}
\end{table*}

% table 4
\begin{table*}
\centering
\caption{Continued.
}
\label{tab5}
    \begin{tabular}{cccccccc} % four columns, alignment for each
    \hline
    ID & R.A. & Dec. & $z_\mathrm{spec}$ & Line & Reference & M$_\star$ [$10^{10}$ M$_\odot$] & SFR [M$_\star$yr$^{-1}$]\\
    \hline
    71 & 11:40:55.18 & $-$26:28:42.0 & 2.1630 & H$\alpha$  & P23 & $3.25\pm1.12$ & $3.27\pm3.20$\\
    72 & 11:40:36.84 & $-$26:28:34.1 & ---    & ---        & --- & $0.44\pm0.19$ & $13.54\pm4.83$\\
    ...\\
    \hline
\end{tabular}
\end{table*}

\end{comment}

\newpage

% table 4
\begin{table*}
\centering
\caption{
Same as Table~\ref{tab1} but for all 84 HAEs updated by this work.
References are written in the following abbreviations (S24: this work, P23: \citealt{Perez-Martinez2023}, J21: \citealt{Jin2021}, S14: \citealt{Shimakawa2014}, C05: \citealt{Croft2005}, K04: \citealt{Kurk2004b}, K00: \citealt{Kurk2000}).
}
\label{tab4}
    \begin{tabular}{cccccccc} % four columns, alignment for each
    \hline
    ID & R.A. & Dec. & $z_\mathrm{spec}$ & Line & Reference & M$_\star$ [$10^{10}$ M$_\odot$] & SFR [M$_\star$yr$^{-1}$]\\
    \hline
    2  & 11:40:55.20 & $-$26:30:42.7 & 2.1447 & H$\alpha$  & P23 & $0.38\pm0.12$ & $15.71\pm4.30$\\
    4  & 11:40:59.81 & $-$26:30:42.6 & 2.1615 & H$\alpha$  & P23 & $1.09\pm0.18$ & $66.79\pm12.60$\\
    5  & 11:40:57.80 & $-$26:30:48.1 & 2.1635 & H$\alpha$  & P23 & $13.19\pm1.14$ & $107.3\pm5.4$\\
    7  & 11:40:59.61 & $-$26:30:39.1 & 2.1645 & H$\alpha$  & P23 & $2.90\pm0.69$ & $82.51\pm12.94$\\
    8  & 11:40:54.28 & $-$26:30:30.6 & 2.1356 & H$\alpha$  & S14 & $0.29\pm0.08$ & $14.67\pm3.62$\\
    9  & 11:40:38.16 & $-$26:30:24.3 & 2.1437 & H$\alpha$  & S14 & $4.11\pm0.91$ & $38.44\pm6.27$\\
    10 & 11:40:58.86 & $-$26:30:22.2 & ---    & ---        & --- & $0.47\pm0.15$ & $19.36\pm4.17$\\
    11 & 11:40:39.43 & $-$26:30:24.8 & 2.1627 & H$\alpha$  & P23 & $1.01\pm0.17$ & $44.74\pm2.24$\\
    12 & 11:40:37.43 & $-$26:30:21.2 & ---    & ---        & --- & $1.46\pm0.55$ & $55.83\pm21.97$\\
    13 & 11:40:58.73 & $-$26:30:22.5 & 2.1650 & H$\alpha$  & P23 & $5.05\pm0.48$ & $74.26\pm26.26$\\
    14 & 11:40:37.34 & $-$26:30:17.3 & 2.1684 & H$\alpha$  & P23 & $19.52\pm6.70$ & $6.92\pm4.67$\\
    15 & 11:40:53.22 & $-$26:30:05.4 & ---    & ---        & --- & $9.43\pm0.47$ & $141.3\pm25.8$\\
    16 & 11:40:52.61 & $-$26:30:00.8 & 2.1577 & H$\alpha$  & P23 & $0.82\pm0.10$ & $38.76\pm2.86$\\
    17 & 11:40:53.67 & $-$26:30:05.8 & 2.1617 & CO         & J21 & $1.28\pm0.19$ & $56.71\pm2.84$\\
    18 & 11:40:40.09 & $-$26:29:47.4 & 2.1609 & H$\alpha$  & P23 & $0.72\pm0.23$ & $22.04\pm5.21$\\
    19 & 11:40:40.11 & $-$26:29:46.6 & ---    & ---        & --- & $0.55\pm0.11$ & $26.19\pm2.75$\\
    21 & 11:40:51.56 & $-$26:29:45.7 & 2.1575 & H$\alpha$  & P23 & $1.06\pm0.26$ & $45.78\pm12.77$\\
    22 & 11:40:47.46 & $-$26:29:41.2 & ---    & ---        & --- & $2.26\pm0.99$ & $35.14\pm17.34$\\
    23 & 11:40:43.42 & $-$26:29:37.3 & 2.1463 & H$\alpha$  & S14 & $0.51\pm0.06$ & $25.76\pm1.50$\\
    24 & 11:40:46.86 & $-$26:29:36.9 & ---    & ---        & --- & $0.16\pm0.05$ & $6.43\pm2.04$\\
    25 & 11:40:57.38 & $-$26:29:37.5 & 2.1659 & H$\alpha$  & P23 & $1.26\pm0.14$ & $59.09\pm2.95$\\
    26 & 11:40:57.64 & $-$26:29:35.4 & ---    & ---        & --- & $0.78\pm0.23$ & $27.95\pm6.33$\\
    27 & 11:40:57.81 & $-$26:29:35.7 & 2.1701 & H$\alpha$  & S14 & $1.90\pm0.24$ & $61.30\pm4.76$\\
    28 & 11:40:50.70 & $-$26:29:33.6 & 2.1532 & CO         & J21 & $7.61\pm3.29$ & $4.82\pm4.70$\\
    29 & 11:40:57.91 & $-$26:29:36.3 & 2.1703 & H$\alpha$  & P23 & $7.46\pm5.65$ & $6.18\pm4.79$\\
    30 & 11:40:51.27 & $-$26:29:38.6 & 2.1513 & H$\alpha$  & P23 & $8.64\pm0.43$ & $71.15\pm3.56$\\
    32 & 11:40:43.13 & $-$26:29:24.3 & ---    & ---        & --- & $0.49\pm0.07$ & $21.07\pm2.71$\\
    33 & 11:40:50.58 & $-$26:29:21.2 & ---    & ---        & --- & $0.46\pm0.12$ & $20.70\pm3.70$\\
    34 & 11:40:49.81 & $-$26:29:22.2 & ---    & ---        & --- & $0.94\pm0.38$ & $36.66\pm13.23$\\
    35 & 11:40:46.13 & $-$26:29:24.7 & 2.1551 & H$\alpha$  & P23 & $2.06\pm0.18$ & $102.6\pm5.1$\\
    36 & 11:40:53.72 & $-$26:29:19.8 & ---    & ---        & --- & $0.40\pm0.10$ & $17.49\pm2.99$\\
    37 & 11:40:46.97 & $-$26:29:19.2 & ---    & ---        & --- & $1.81\pm0.59$ & $41.31\pm13.43$\\
    38 & 11:40:50.19 & $-$26:29:20.9 & 2.1543 & H$\alpha$  & P23 & $0.48\pm0.07$ & $24.98\pm2.33$\\
    39 & 11:40:46.29 & $-$26:29:24.3 & ---    & ---        & --- & $5.14\pm1.29$ & $97.94\pm18.64$\\
    40 & 11:40:44.48 & $-$26:29:20.6 & 2.1620 & Ly$\alpha$ & C05 & $9.19\pm2.00$ & $5.33\pm3.43$\\
    41 & 11:40:45.77 & $-$26:29:18.8 & ---    & ---        & --- & $0.63\pm0.10$ & $29.98\pm2.20$\\
    42 & 11:40:48.69 & $-$26:29:16.4 & ---    & ---        & --- & $0.80\pm0.26$ & $31.46\pm5.16$\\
    43 & 11:40:36.95 & $-$26:29:16.8 & ---    & ---        & --- & $0.48\pm0.06$ & $23.79\pm1.54$\\
    44 & 11:40:37.78 & $-$26:29:12.4 & 2.1574 & H$\alpha$  & P23 & $4.78\pm0.68$ & $171.5\pm8.6$\\
    45 & 11:40:48.09 & $-$26:29:11.3 & 2.1415 & H$\alpha$  & S14 & $0.39\pm0.07$ & $13.14\pm1.48$\\
    46 & 11:40:45.98 & $-$26:29:16.7 & 2.1557 & H$\alpha$  & P23 & $2.10\pm3.24$ & $7.87\pm7.66$\\
    48 & 11:40:46.67 & $-$26:29:10.3 & 2.1663 & H$\alpha$  & P23 & $18.32\pm2.16$ & $3.53\pm0.45$\\
    49 & 11:40:49.39 & $-$26:29:09.1 & 2.1661 & H$\alpha$  & P23 & $0.29\pm0.09$ & $14.01\pm3.01$\\
    51 & 11:40:44.09 & $-$26:29:05.6 & ---    & ---        & --- & $0.27\pm0.15$ & $6.78\pm3.50$\\
    53 & 11:40:52.99 & $-$26:29:04.5 & ---    & ---        & --- & $0.67\pm0.36$ & $15.59\pm9.43$\\
    54 & 11:40:46.07 & $-$26:29:11.3 & 2.1480 & H$\alpha$  & S14 & $16.04\pm0.80$ & $130.3\pm9.8$\\
    55 & 11:40:44.25 & $-$26:29:07.0 & 2.1583 & H\&K       & S24 & $23.23\pm2.92$ & $2.54\pm1.23$\\
    56 & 11:40:48.35 & $-$26:29:05.0 & ---    & ---        & --- & $0.58\pm0.30$ & $16.69\pm7.08$\\
    57 & 11:40:45.54 & $-$26:29:02.2 & 2.1523 & H$\alpha$  & K04 & $0.25\pm0.04$ & $12.17\pm1.05$\\
    58 & 11:40:47.95 & $-$26:29:06.1 & 2.1568 & H$\alpha$  & P23 & $6.79\pm1.13$ & $41.34\pm15.57$\\
    59 & 11:40:47.30 & $-$26:28:55.5 & ---    & ---        & --- & $0.21\pm0.07$ & $8.40\pm2.52$\\
    60 & 11:40:41.68 & $-$26:28:54.2 & 2.1634 & H$\alpha$  & S14 & $1.42\pm0.32$ & $62.33\pm11.38$\\
    61 & 11:41:00.14 & $-$26:28:56.3 & 2.1665 & H$\alpha$  & P23 & $5.27\pm0.46$ & $64.85\pm10.29$\\
    62 & 11:41:03.28 & $-$26:28:51.0 & 2.1493 & H$\alpha$  & P23 & $0.86\pm0.26$ & $35.74\pm10.44$\\
    63 & 11:40:39.71 & $-$26:28:49.0 & ---    & ---        & --- & $0.31\pm0.08$ & $12.34\pm2.58$\\
    64 & 11:40:49.49 & $-$26:28:48.6 & ---    & ---        & --- & $0.55\pm0.17$ & $22.63\pm5.90$\\
    65 & 11:40:47.98 & $-$26:28:49.6 & 2.1628 & H$\alpha$  & P23 & $0.92\pm0.31$ & $30.74\pm7.67$\\
    66 & 11:40:54.81 & $-$26:28:45.1 & ---    & ---        & --- & $0.63\pm0.16$ & $23.04\pm1.72$\\
    67 & 11:40:44.14 & $-$26:28:44.0 & 2.1634 & H$\alpha$  & S14 & $0.48\pm0.07$ & $21.34\pm2.43$\\
    68 & 11:40:39.73 & $-$26:28:45.2 & 2.1620 & Ly$\alpha$ & C05 & $14.72\pm4.13$ & $35.88\pm30.98$\\
    69 & 11:40:51.67 & $-$26:28:41.3 & ---    & ---        & --- & $1.04\pm0.29$ & $16.54\pm4.45$\\
    70 & 11:40:46.85 & $-$26:28:41.1 & 2.1636 & H$\alpha$  & P23 & $0.27\pm0.07$ & $12.43\pm3.29$\\
    \hline
\end{tabular}
\end{table*}

% table 4
\begin{table*}
\centering
\caption{Continued.
}
\label{tab5}
    \begin{tabular}{cccccccc} % four columns, alignment for each
    \hline
    ID & R.A. & Dec. & $z_\mathrm{spec}$ & Line & Reference & M$_\star$ [$10^{10}$ M$_\odot$] & SFR [M$_\star$yr$^{-1}$]\\
    \hline
    71 & 11:40:55.18 & $-$26:28:42.0 & 2.1630 & H$\alpha$  & P23 & $3.25\pm1.12$ & $3.27\pm3.20$\\
    72 & 11:40:36.84 & $-$26:28:34.1 & ---    & ---        & --- & $0.44\pm0.19$ & $13.54\pm4.83$\\
    73 & 11:40:48.36 & $-$26:29:08.7 & 2.1618 & CO         & J21 & $261.7\pm39.4$ & $542.4\pm81.1$\\
    74 & 11:40:44.86 & $-$26:28:40.6 & 2.1435 & H$\alpha$  & S14 & $0.46\pm0.19$ & $15.39\pm4.54$\\
    75 & 11:40:57.10 & $-$26:28:27.9 & 2.1576 & H$\alpha$  & P23 & $0.87\pm0.13$ & $49.19\pm4.76$\\
    76 & 11:40:44.75 & $-$26:28:25.9 & ---    & ---        & --- & $0.42\pm0.12$ & $14.91\pm1.92$\\
    77 & 11:40:55.29 & $-$26:28:23.8 & ---    & Ly$\alpha$ & K00 & $3.22\pm1.42$ & $5.73\pm4.23$\\
    79 & 11:40:56.36 & $-$26:28:23.7 & 2.1892 & CO         & J21 & $0.88\pm0.29$ & $36.67\pm11.72$\\
    80 & 11:40:54.56 & $-$26:28:23.7 & 2.1606 & H$\alpha$  & P23 & $4.74\pm0.78$ & $185.2\pm9.3$\\
    82 & 11:40:45.77 & $-$26:28:12.5 & ---    & ---        & --- & $3.91\pm0.67$ & $25.03\pm41.57$\\
    83 & 11:40:45.50 & $-$26:28:10.2 & ---    & ---        & --- & $2.42\pm1.11$ & $5.85\pm6.14$\\
    84 & 11:40:42.87 & $-$26:28:07.2 & 2.1623 & H$\alpha$  & P23 & $0.83\pm0.13$ & $34.94\pm1.93$\\
    85 & 11:40:36.84 & $-$26:28:03.1 & ---    & Ly$\alpha$ & K00 & $0.42\pm0.10$ & $19.55\pm3.63$\\
    87 & 11:40:49.99 & $-$26:27:54.8 & ---    & ---        & --- & $0.19\pm0.07$ & $7.26\pm1.96$\\
    88 & 11:41:02.44 & $-$26:27:49.7 & ---    & ---        & --- & $0.32\pm0.08$ & $12.82\pm2.31$\\
    89 & 11:40:47.37 & $-$26:27:59.1 & ---    & ---        & --- & $4.95\pm0.39$ & $0.42\pm1.77$\\
    90 & 11:40:55.67 & $-$26:27:23.6 & ---    & ---        & --- & $0.46\pm0.20$ & $10.83\pm4.65$\\
    92 & 11:40:59.18 & $-$26:27:56.3 & 2.1471 & H$\alpha$  & P23 & $0.78\pm0.22$ & $40.69\pm9.80$\\
    93 & 11:40:54.75 & $-$26:28:03.2 & 2.1516 & H$\alpha$  & P23 & $2.45\pm1.09$ & $73.61\pm25.58$\\
    94 & 11:41:01.24 & $-$26:27:41.7 & ---    & ---        & --- & $0.40\pm0.20$ & $8.48\pm3.50$\\
    95 & 11:41:02.39 & $-$26:27:45.1 & 2.1510 & H$\alpha$  & P23 & $12.78\pm16.37$ & $65.23\pm29.66$\\
    96 & 11:40:44.41 & $-$26:27:43.0 & ---    & ---        & --- & $0.15\pm0.05$ & $7.42\pm2.16$\\
    \hline
\end{tabular}
\end{table*}

%%%%%%%%%%%%%%%%%%%%%%%%%%%%%%%%%%%%%%%%%%%%%%%%%%

% Don't change these lines
\bsp	% typesetting comment
\label{lastpage}
\end{document}